\documentclass[14pt, preprint2]{aastex6}
\usepackage{bm, graphicx, subfigure, amsmath, morefloats, mathtools, amsfonts}
\hypersetup{allcolors = [rgb]{0., 0., 0.5}} 

\usepackage{longtable}
\newcommand{\project}[1]{\textsl{#1}}
 
\newcommand{\tc}{\project{The~Cannon}} 
\newcommand{\apogee}{\project{\textsc{apogee}}}
\newcommand{\apokasc}{\project{\textsc{apokasc}}}
\newcommand{\aspcap}{\project{\textsc{aspcap}}}

\newcommand{\kepler}{\project{Kepler}}
\newcommand{\sloanv}{\project{Sloan V}}
\newcommand{\mwm}{\project{Milky Way Mapper}}

\newcommand{\galah}{\project{\textsc{galah}}}

\newcommand{\documentname}

\newcommand{\teff}{\mbox{$T_{\rm eff}$}}

\newcommand{\feh}{\mbox{$\rm [Fe/H]$}}

\newcommand{\alphafe}{\mbox{$\rm [\alpha/Fe]$}}

\newcommand{\logg}{\mbox{$\log g$}}

\newcommand{\rgal}{\mbox{$R_{\text{GAL}}$}}

\newcommand{\set}[1]{\bm{#1}}
\newcommand{\starlabel}{\ell}
\newcommand{\starlabelvec}{\set{\starlabel}}

\newcommand{\xone}{\ensuremath{x_n}}
\newcommand{\xtwo}{\ensuremath{x_{n'}}}


\begin{document}






\title{In the Galactic disk, stellar [Fe/H] and age predict orbits and precise [X/Fe]}

\author{M.K.~Ness\altaffilmark{1,2},  K.V. Johnston,\altaffilmark{1}, K.  Blancato\altaffilmark{1},  H-W.~Rix\altaffilmark{3},
 A. Beane\altaffilmark{4}, J.C Bird\altaffilmark{5}, K. Hawkins\altaffilmark{6}}

\altaffiltext{1}{Department of Astronomy, Columbia University, Pupin Physics Laboratories, New York, NY 10027, USA}\
\altaffiltext{2}{Center for Computational Astrophysics, Flatiron Institute, 162 Fifth Avenue, New York, NY 10010, USA}
\altaffiltext{3}{Max-Planck-Institut f\"ur Astronomie, K\"onigstuhl 17, D-69117 Heidelberg, Germany}
\altaffiltext{4}{Department of Physics \& Astronomy, University of Pennsylvania, 209 South 33rd Street, Philadelphia, PA 19104, USA}
\altaffiltext{5}{Department of Physics and Astronomy, Vanderbilt University, 6301 Stevenson Center, Nashville, TN, 37235, USA}
\altaffiltext{6}{Department of Astronomy, The University of Texas at Austin, 2515 Speedway Boulevard, Austin, TX 78712, USA}

\email{melissa.ness@columbia.edu}


\begin{abstract} 

We explore the  structure of the element abundance--age--orbit distribution of the stars in the Milky Way's low-$\alpha$ disk, by (re-)deriving precise \feh, [X/Fe] and ages, along with orbits, for red clump stars from the \apogee\ survey. There has been a long-standing theoretical expectation and observational evidence that metallicity (\feh) and age are informative about a star's orbit, e.g. about its angular momentum and the  corresponding mean Galactocentric distance or its vertical motion. Indeed, our analysis of the \apogee\ data confirms that \feh\ or age alone can predict the stars' orbits far less well than the combination of the two. Remarkably, we find and show explicitly, that for known \feh\ and age, the other abundances [X/Fe] of Galactic disk stars can be predicted well (on average to 0.02 dex) across a wide range of Galactocentric radii, and therefore provide little additional information, e.g. for predicting their orbit. While the age-abundance space for metal poor stars and potentially for stars near the Galactic center is rich or complex, for the bulk of the Galaxy's low-$\alpha$ disk it is simple: \feh\ and age contain most information, unless [X/Fe] can be measured to 0.02, or better. Consequently, we do not have the precision with current  (and likely near-future) data to assign stars to their individual (coeval) birth clusters, from which the disk is presumably formed. We can, however, place strong constraints on future models of galactic evolution, chemical enrichment and mixing. 

\end{abstract}

\section{Introduction}

In the pursuit of using stars to understand the processes that have formed and evolved the Milky Way disk, many millions of low and medium resolution spectra have been obtained \citep[e.g.][]{Steinmetz2006, Yanny2009, Gilmore2012, Newberg2012,  deSilva2015, Majewski2017}. The acquisition rate of stellar spectral data-intake will increase into the next decade \citep[e.g.][]{Kollmeier2017, Bonifacio2016, deJong2019, deSilva2015, PFS2018, Newberg2012, C2014}. Stellar abundances are expected to encode the temporal enrichment of the disk. An ensemble of abundance measurements, for vast stellar samples, over large spatial extents, is therefore critical to constrain the diversity and characteristics of nucleosynthetic channels. 
From a theoretical standpoint,  it is also expected that given precise enough chemical abundance measurements and sufficient data, we might also be able to reconstruct the original birth sites of stars in the Galactic disk. That is, to assign them to the individual clusters from which they formed  \citep{Arm2018, Krumholz2018, BH2010}.  However, the empirical information content and diagnostic power of the multitude of element abundances being assembled, in terms of understanding the disk's formation, has to date, been poorly constrained \citep[see however][]{Ting2012}

At this point, we have established, from the large number of large Milky Way stellar surveys and subsequent studies, a global set of descriptive properties of the Galactic disk. These pertain to gradients in metallicity (\feh), age and $\alpha$-element enhancements. Stars in the disk show a vertical abundance gradient \citep{Katie2014}, of about -0.24 dex/kpc \citep[e.g.][]{Duong2018} which increases to -0.45 dex/kpc in the bulge \citep[e.g.][]{Ness2013a}.  Radially, the young stars confined to the plane of the Milky Way disk show a weak gradient in their \feh, of {--0.075 dex/kpc} \citep[e.g.][]{G2014, Frankel2018}. This gradient, and the small dispersion around it among young stars, presumably reflects an initial birth relation between the level of chemical enrichment of the star-forming gas and its radii. Such a correlation would be weakened by any subsequent evolutionary processes that have moved stars from their birth places over time \citep[e.g.][]{Selwood2002, Roskar2008, Minchev2011}. This is directly seen in global models of the age-\feh\ relation of the Galactic disk \citep{Frankel2018}, as stars in the Milky Way suffer significant dynamical `memory loss' over time, through radial migration. That said, the weak trends seen between orbits and \feh\ at all ages indicate there is some level of memory retention linking orbits to chemical compositions \citep{Beane2018}.

In addition to the metallicity gradient seen in the disk, bulk element enhancement trends, like [$\alpha$/Fe], are dramatically informative as to location in the disk.  The  mapping of [$\alpha$/Fe] across the disk, from Galactic radii 4 -- 16 kpc, shows that stars with high $\alpha$-enhancement are concentrated to the inner galaxy and stars with low $\alpha$-enhancement are concentrated to the outer galaxy \citep[e.g.][]{Bensby2012}. Both the high and low $\alpha$-sequence appear near the Sun. The fraction of  $\alpha$-enhanced stars also increases at larger heights from the plane, at a given radius \citep{Bovy2012, Hayden2015, Nidever2014}. Broadly, the $\alpha$-enhancement trends seen across the Galaxy imply a faster rate of star formation in the inner Galaxy compared to the outer Galaxy. These trends reflect the convolution of many physical processes, including the star formation history, disk growth and heating and migratory processes during the disk's evolution. 
The two $\alpha$-enhancement sequences as described visually in the \feh-[$\alpha$/Fe] plane, have been demonstrated to be empirically dynamically different \citep[e.g][]{Ted2019b, Gandhi2019} and explained theoretically as having different Galactic formation origins \citep[e.g.][]{Ted2019a, Clarke2019}.

The \apogee\ ($\approx$ 500,000 stars) survey \citep{Majewski2017} and its successor, the upcoming ($>$ 5 million star) \sloanv\ \mwm\ survey \citep{Kollmeier2017} are benchmark Galactic archaeology programs. These are providing the data to trace the chemo-dynamical structure of the Galaxy over a large spatial extent, by observing hundreds of thousands, and from 2020, with \mwm, millions, of bright giant tracers. The \galah\ survey was engineered to be the ultimate `chemical tagging' experiment, observing an ensemble of elements from a larger multitude of nucleosynthetic families, primarily for nearby main sequence stars, in large enough number to reconstruct individual cluster birth sites using abundances measured for individual stars. Critically, from current surveys (and for future surveys), we now have, from  data-driven modeling using benchmark stars with precision asteroseismic ages, access not only to abundances, but to ages for large numbers of stars across the Milky Way disk \citep[e.g.][]{Martig2016, Ness2016,  Ho2017b, Leung2019, Bovy2019}.  We can see from the \apogee\ survey for example, evidence for the inside-out formation of the Milky Way from these ages \citep{Ness2018}. Age is the fundamental variable of temporal evolution, and the key tracer around which all other variables can be pivoted in the pursuit of using the ensemble of stellar measurements to understand our Galaxy.

In this paper, we seek to answer the following fundamental questions about the information content in the data: (i) what does each abundance tell us about stellar age, (ii) is there additional information content in each abundance at a given age and (iii) how do the present day orbits of stars of different mean abundances vary, as a function of age at fixed \feh?  

We use orbit actions to quantify the stellar orbital properties \citep{BT2008}. Using \textit{Gaia} parallaxes and proper motions, we can access these actions. The three actions, of $J_R$, $J_z$ and $J_\phi$ are each expressions of the radial excursion around the guiding radius, the vertical extent of the orbit above its orbital plane and its guiding, or average radius, respectively.

We use approximately 20,000 red clump stars identified in the catalogue of \citet{Ting2018}, with abundances that have been corrected for known systematics, from the  \apogee\ survey. From this set of stars, we isolate the $\approx$ 15,000 stars in the low-$\alpha$ sequence. We examine only the low-$\alpha$ stars as we wish to exclude stars in the high-$\alpha$ sequence with a probable separate origin of formation \citep[see][]{Gandhi2019,Ted2019a, Clarke2019}. The \apogee\ red clump sample extend across a wide radial extent (4-16 kpc) and from these spectra we measure a multitude of elements.  From this set of stars, we derive a set of precision chemical abundances, ages precise to $\lesssim$ 1.6 Gyr and orbital actions precise to $<$ 10 percent. We determine 19 abundances using a data-driven modeling training on \aspcap\ measurements and correcting for systematics introduced by the varying line spread function across the spectrograph \citep{Ness2018}. 

We have a particularly valuable set of $\approx$ 1100 local red clump stars in this sample, from the APOKASC catalogue  \citep{Pins2018}. These stars have masses and subsequently ages determined from their internal oscillations via asteroseismology, from their Kepler power spectra. We use a subset of this sample to investigate the relationship between element abundances [X/Fe] and age, and compare our results to very high precision analyses from R=115,000 spectra of main sequence and turn-off solar-twin stars \citep{Bedell2018}. We also use the \apokasc\ stars as a reference set, to determine ages for the full red clump low-$\alpha$ sample, using data-driven modeling. 

To examine the intrinsic information content in each abundance as a function of age, we measure the intrinsic abundance dispersion around the age-abundance relationships, for each element. That is, any additional dispersion in the abundances as a function of age that can not be explained by the age and abundance measurement errors. Such intrinsic dispersions directly quantify to what level each element may differentiate different sites of star formation at a given time, which has been theoretically expected but not empirically demonstrated \citep[e.g.][]{Krumholz2018, A2018}. Furthermore, the intrinsic dispersion indicates the precision at which each element must be measured in order to render that element informative beyond being a label of age, for any individual star. We also examine the trends between stellar abundances and orbital action labels.  We consider populations of stars, combined using their abundance similarity, to demonstrate that stars separate out dynamically and temporally at fixed \feh, and the separation is dissimilar between groups of old and groups of young stars. This population analysis is a practical demonstration that abundances indicate age, and orbital properties of the disk architecture. 

Finally, we discuss the implications of our findings, which demonstrate the power of abundances to indicate ages and that the  majority of information about a star e.g. linking to orbits, is captured in its age and \feh, for the low-$\alpha$ disk.  We conclude citing the prospects for the coming era of multi-million star surveys, where combining measurements of stars will tap into the orbital sub-structure of the disk, opening up opportunities in Galactic archeology across a breadth of spatial scales.

\section{Data}

Our data is sourced from the \apogee\ DR14 \citep{Majewski2017} catalogue, the second \apokasc\ data release of stellar ages \citep{Pins2018} and the \textit{Gaia} DR2 \citep{Gaia2018} data release. For the \apogee\ stars, we use the $\approx$ 20,000 spectroscopically confirmed red clump stars \citep{Hawkins2018}. This sample represents the most pristine red clump selection available with an expected contamination rate of 3 percent \citep{Ting2018}. These stars have precise distances from their fixed red clump absolute magnitudes \citep{Hawkins2017}. Of these stars, we select $\approx$ 15,000 low-alpha sequence stars for our analysis, based on their location in the \feh-\alphafe\ plane. To determine abundances for our red clump stars, we use \tc\ on the the DR14 spectra, with the \aspcap\ abundances and stellar parameters as the training labels, following the prescription and training set of \citet{Ness2018}.  The \apogee\ survey delivers a multitude of element abundances: we learn labels of \feh\ plus 18 additional abundance measurements from the \apogee\ red clump infrared spectra. These elements are C, N (light proton), Na, Al, S, K (light odd-z), Mg, Si, Ca, Ti, O (alpha), Fe, V, Mn, Ni, P, Cr, Co (iron-peak) and  Rb (s-process). The elements Na and Rb have high uncertainties and are not robustly recovered in our data-driven inference and we end up excluding these elements from our analysis. 

\begin{figure}[]
\includegraphics[scale=0.3]{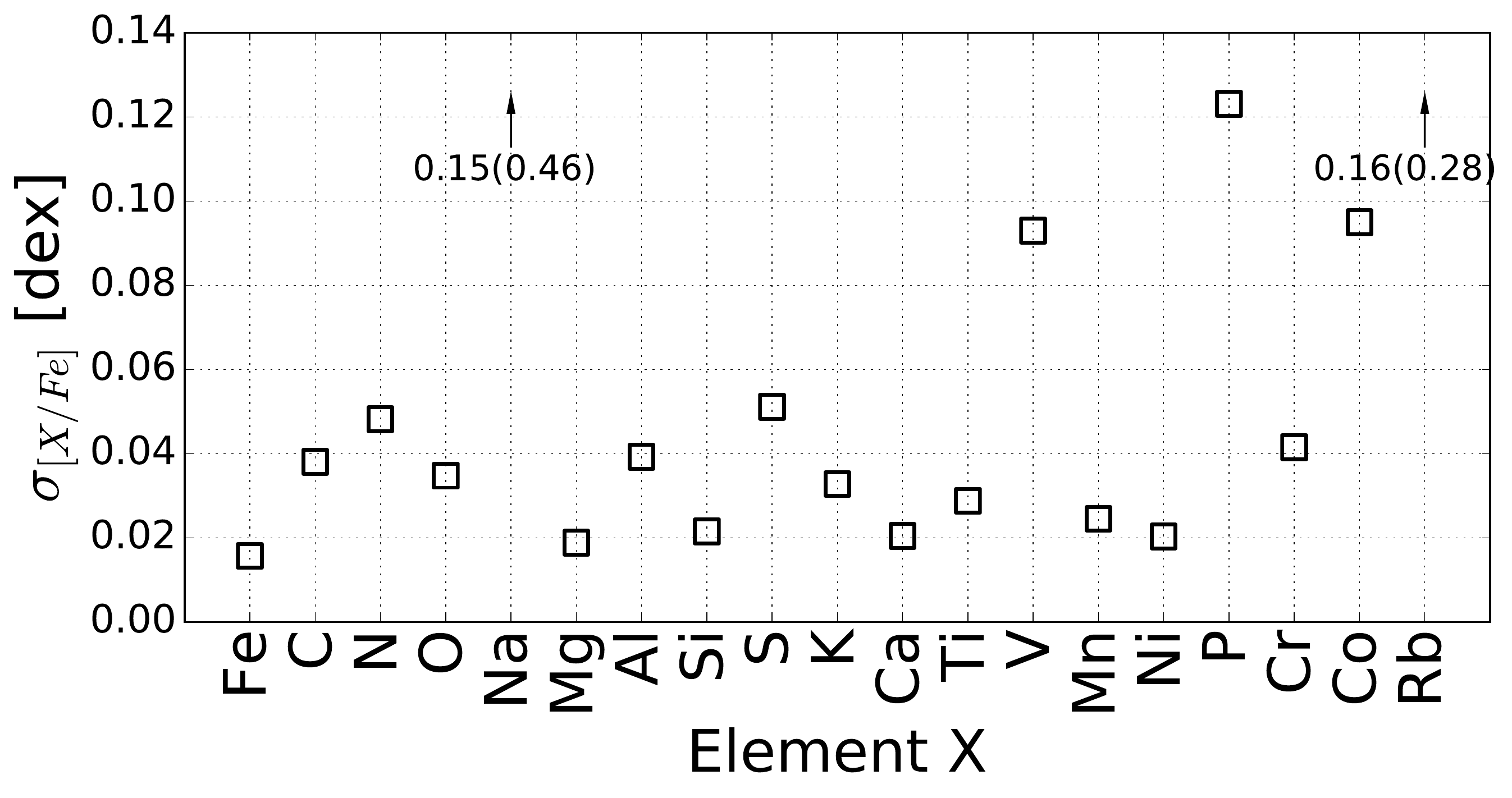} 
\caption{The mean uncertainty for each abundance measurement that we derive from \apogee\ spectra using \tc\ \citep{Ness2015}. These errors are the \textit{rms} sum of the cross-validation error of a subset of $\approx$ 2200 of the stars (that span the narrow red clump parameter space) from the $\approx$ 4500 star training set of \tc\ \citep{Ness2018a}, and the formal errors, as reported from the optimization in \tc. Note the formal errors are typically about 10 percent of the magnitude of the cross-validation errors. The elements Na and Rb fail for a subset of stars in cross-validation and the bracketed errors indicate what the error is including these failed stars (if we exclude the failures at cross-validation the errors are 0.15 and 0.16 dex respectively, as indicated). Na and Rb are largely excluded from our analysis as a consequence of their cross-validation failure for a fraction of the stars.} 
\label{fig:elemerr}
\end{figure}

We take care with our abundance derivations to remove the remaining systematic abundance trends that are apparent in the abundance correlations seen with fiber number \citep[as described in][]{Ness2018}. These trends are significant and if not accounted for, may propagate into problematic systematic effects when abundances are used in concert for analyses. We determine the error on our abundances using a root mean square (\textit{rms}) sum of the cross-validation error from \tc's inference and the formal uncertainty of \tc's optimizer. For the cross-validation errors, we generate 20 models, for each model excluding 5 percent of the data. We subsequently test how well the abundances are recovered for the excluded stars, determining the \textit{rms} difference of the true (\aspcap) minus the inferred values from \tc. Our analysis uses only the red clump stars, so we restrict our cross-validation test to the $\approx$ 2200 training stars within the red clump parameter space of \teff = 4820 K $\pm$ 200 K, \logg\ = 2.45 $\pm$ 0.5 and \feh\ = 0 $\pm$ 0.5 dex. This additional restriction to the overall parameter space of the training set leads to only a marginal or negligible improvement in the \textit{rms} uncertainty for the element abundances but a marked improvement in the \feh\ precision, which decreases from 0.022 dex to 0.015 dex. The mean uncertainty of these abundance measurements overall is $<$ 0.03 dex (see Figure \ref{fig:elemerr}).  In our cross-validation analysis, we find that we fail to recover the Na and Rb abundances for a subset of the stars. We therefore exclude these two elements from the majority of our analysis.

\begin{figure*}[]
\centering
\includegraphics[scale=0.5]{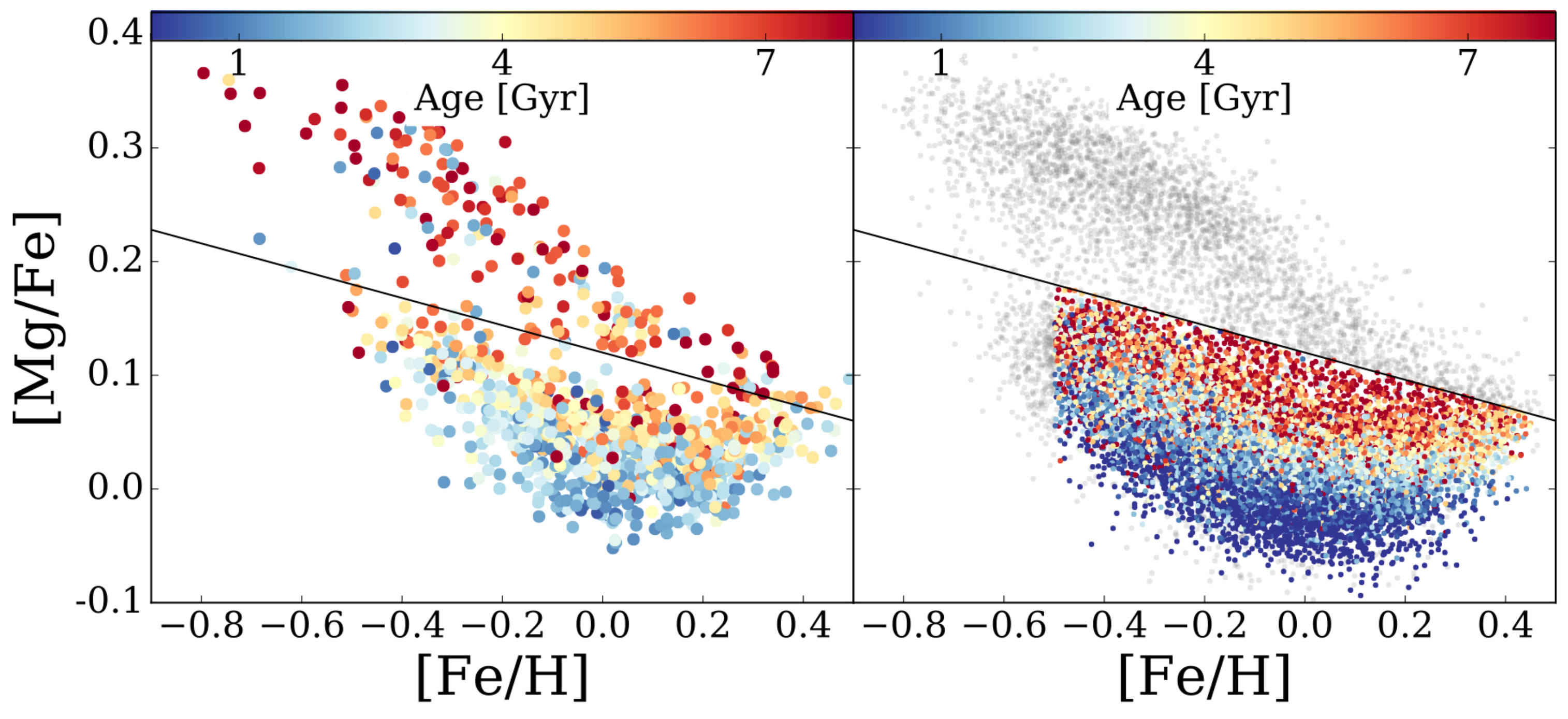} 
\caption{The stellar ages shown in the \feh-[Mg/Fe] plane. At left, the $\approx$ 1100 \apokasc\ red clump sample of stars with asteroseismic ages \citep{Pins2018}. At right, the larger sample of \apogee\ red clump stars, for which we derive ages, which are shown, using a model built from the low-$\alpha$ asteroseismic stars.  The age-scale is provided at the top of each panel.  The $\approx$ 800 red clump stars in the \apokasc\ sample that meet our quality cuts at left are  described in Section 3.1 (with 600 stars in the low-$\alpha$ sample). The full 20,000 red clump sample of stars in \apogee\ shown in the panel at right have membership determined from \citep{Ting2018}. The subset of $\approx$ 15,000 stars for which we determine ages in the right hand panel is described in Section 3.2. The age range for the right hand panel is broader than the left as we place no prior on our data-driven age modeling and our full set of red clump derived ages span a larger range than the \apokasc\ reference sample that we use to build our model (including to negative values). For our analysis, we consider stars in the low-$\alpha$ sequence only, selected as below the (ad hoc) line drawn. This line is intended to exclude stars with a probable separate, different formation history to those of the low-$\alpha$ stars that describe the chemical abundance space of the subsequent inside-out formed Milky Way disk. }
\label{fig:2panel_first}
\end{figure*}

We determine spectroscopic ages for the red clump stars following \citep{Ness2016}. However, instead of modeling the spectral features to determine age labels, we use the precision abundances directly, as our features to learn the relationship between abundances (features) and age (labels). Our training set in this case is comprised of the red clump stars in the second data release of the \apokasc\ catalogue of \citep{Pins2018}, for which there are $\approx$ 1100 red clump stars in total with asteroseismic ages. We implement a number of quality cuts described in Section 3.2. This delivers ages for the full red clump sample that are precise to $\lesssim$ 1.6 Gyrs.

We calculate actions for the \apogee\ stars using their radial velocities (from \apogee), proper motions from \textit{Gaia} DR2 and distances assuming their red clump membership \citep{Ting2018}, using the Galpy code \citep{Bovy2015}. We use solar reference values of 220kms$^{-1}$ and R$_{GAL}$ = 8 kpc and adopt the potential of MWPotential14. The radial velocity precision for \apogee\ is $<$ 300m$s^{-1}$. Under the epicyclic approximation, the total error on the actions can be approximated as the \textit{rms} sum of the fractional error of the parameters input to the action calculation and dominated by the distance error, which is typically 5 percent (errors on our proper motion are typically $<$ 1 percent).  This gives us a high fidelity sample of stars with precision measurements on actions that are typically $<$ 8 percent, that cover a large radial extent across the disk, with a multitude of element families with precision abundance measurements of typically $<$ 0.03 dex.

In this work we analyze the low-$\alpha$ stars, our selection of which is shown in Figure \ref{fig:2panel_first}. The left-hand panel shows the asteroseismic sample of stars colored by their asteroseismic stellar ages calculated from their power spectra \citep{Pins2018} and the right hand panel shows the full set of 20,000 \apogee\ red clump stars from which we derive ages for the $\approx$ 15,000 stars in the low-$\alpha$ sequence across the parameter range of the asteroseismic training set, as described in Section 3.2.

We proceed by first examining the small sample of stars with precise and accurate ages, from asteroseismic measurements in the \apokasc\ catalogue. We use $\approx$ 70 of these stars at solar metallicity, \feh\ = 0 $\pm$ 0.05 dex, to examine the abundance-age trends and the mean intrinsic dispersion of each element. We then use a larger sample of $\approx$ 600 of the asteroseismic red clump stars across a wide range of \feh\ to train a data-driven model to label $\approx$ 15,000 \apogee\ red clump stars with ages from their abundances. The $\approx$ 15,000 stars correspond to those stars within the stellar parameter space of the training set with reference ages that we use to build our model. We use this set of 600 stars with asteroseismic ages to measure the intrinsic dispersion of each element in further detail, modeling age and \feh\ together. We finally examine  the abundance-orbital properties of stars using our sample of $\approx$ 15,000 stars across the disk for which we have derived ages.

\section{Abundance-age correlations in the low-$\alpha$ disk} 

\subsection{The local APOKASC asteroseismic sample of stars at solar \feh}

\begin{figure*}[]
\centering
\includegraphics[scale=0.5]{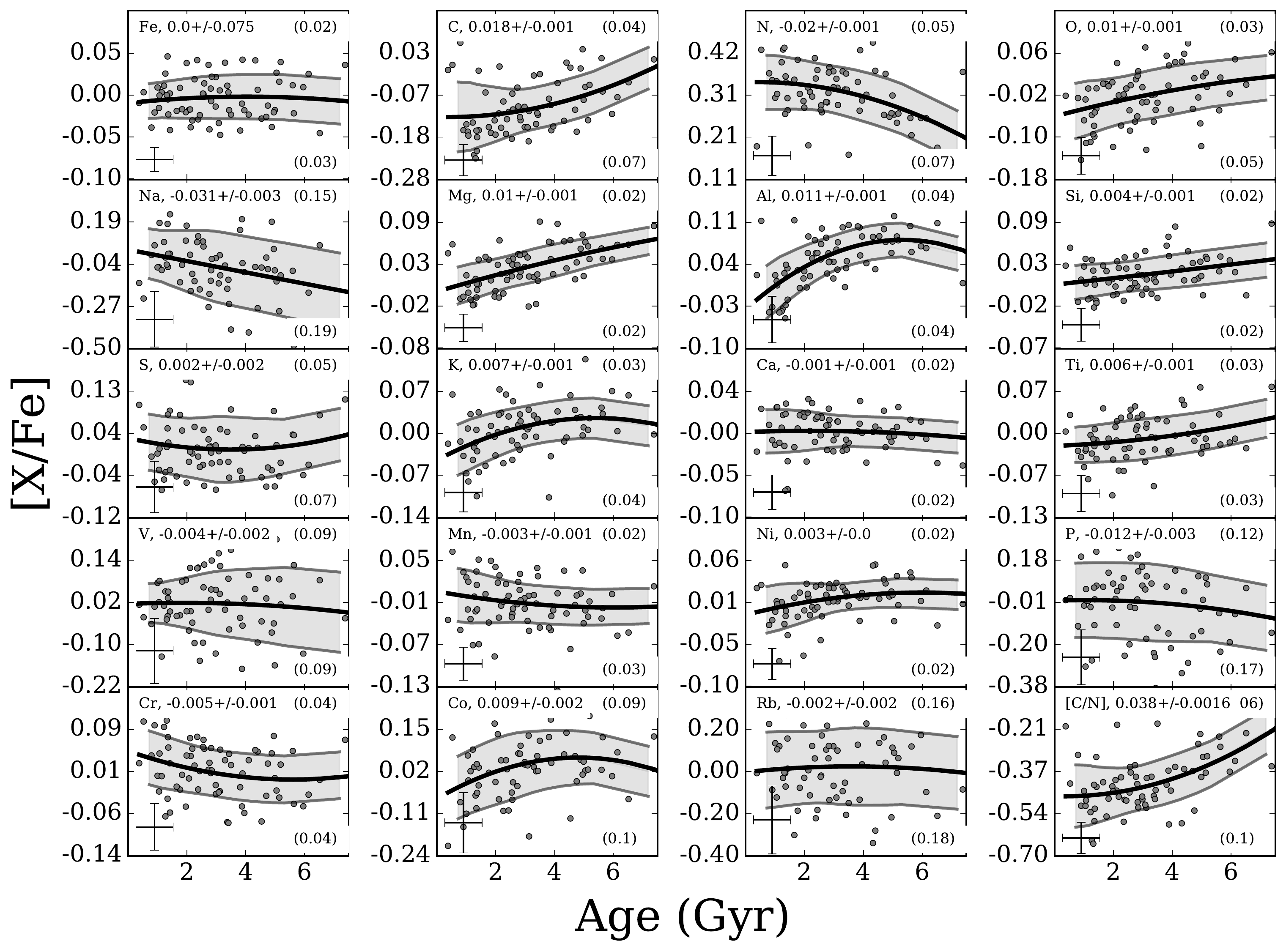}
\caption{This Figure summarizes the small scatter around the  age-abundance relations at fixed age and \feh. Shown are the $\approx$ 70 stars with asteroseismic ages in low-$\alpha$ sequence with \feh\ = 0 $\pm$ 0.05 dex and SNR $>$ 150, showing the age-abundance plane for 19 derived element abundances (we also show [C/N] as this is used as an empirical age indicator i.e.  \citet{Martig2016}). The best fit 2nd-order polynomial line is drawn through the age-abundance plane (in black), and the 1-$\sigma$ dispersion around this line is shown by the shaded grey regions around this central fit. Note that the top left-hand sub-panel indicates the selection around \feh\ = 0 $\pm$ 0.05 dex and the remaining panels are the elements themselves and corresponding line of best fit given this selection. The typical error on the measurement is shown with the error bar at the bottom left of each sub-panel and the error in the y-axis direction is included in brackets in the right-hand corner of each sub-panel. The slope of the best fit straight line is shown, for reference, at the top of each sub-panel, to indicate the scale of the age-abundance correlation for each element (note the y-axes ranges are scaled to best show the data for each element). The bottom right hand bracketed number is the total mean \textit{rms} dispersion of the data around the best fit line ($\sigma_{TOTAL}$).}
\label{fig:ageabund2a}
\end{figure*}

The $\approx$ 600 stars which meet our analysis criteria span a metallicity range of --0.5 $<$ \feh\ $<$ 0.45 dex (with a median \feh\ $\approx$ 0 dex) and an age range of 0.3 $<$ $\tau_{age}$ $<$ 11.7 Gyr (with a median age $\approx$ 3.5 Gyr). We wish to examine the age-abundance correlations of these stars. The left hand panel of Figure \ref{fig:2panel_first} shows clearly that the mean abundance, in this case [Mg/Fe], changes as a function of overall metallicity, \feh. Therefore, to proceed with our investigation, we condition on a single \feh\ value.  From the set of $\approx$ 600 stars with asteroseismic ages that meet our quality cuts, we select the $\approx$ 70 solar metallicity stars, with \feh\ = 0 $\pm$ 0.05 dex. The individual abundances of these stars  versus \feh,  are shown in Figure \ref{fig:ageabund2a}. The best fitting (2nd-order) polynomial line to each age-abundance trend is shown in the thick black line in each of the sub-panels. This line is fit using a least-squares minimization between the line and individual points. For reference, the slope of the best fitting straight line is given in the top of each sub-panel, to indicate the relative magnitude of the gradients for each element (as all y-axes are scaled on a per-element basis). The two shaded grey lines around the line of best fit represent the 1-$\sigma$ dispersion of the data (taking 10 stars per bin) around this line.  

Figure \ref{fig:ageabund2a} shows that the elements C and N, which are associated with mass-dependent evolutionary surface abundance changes in red giants \citep{Martig2016, Masseron2015} have the steepest age-abundance gradients. The elements O and Mg, which more likely reflect temporal chemical enrichment of the disk \citep[e.g.][]{Bensby2017}, are the second most correlated elements with age. Notably, the $\alpha$-elements show a significant variation in their slopes. This is presumably indicative of the different chemical enrichment pathways and subsequent mass dependent yield variations, even within a single nucleosynthetic family of elements \citep[e.g.][]{Weinberg2018, Blancato2019}.

\begin{figure*}[]
\centering
\includegraphics[scale=0.5]{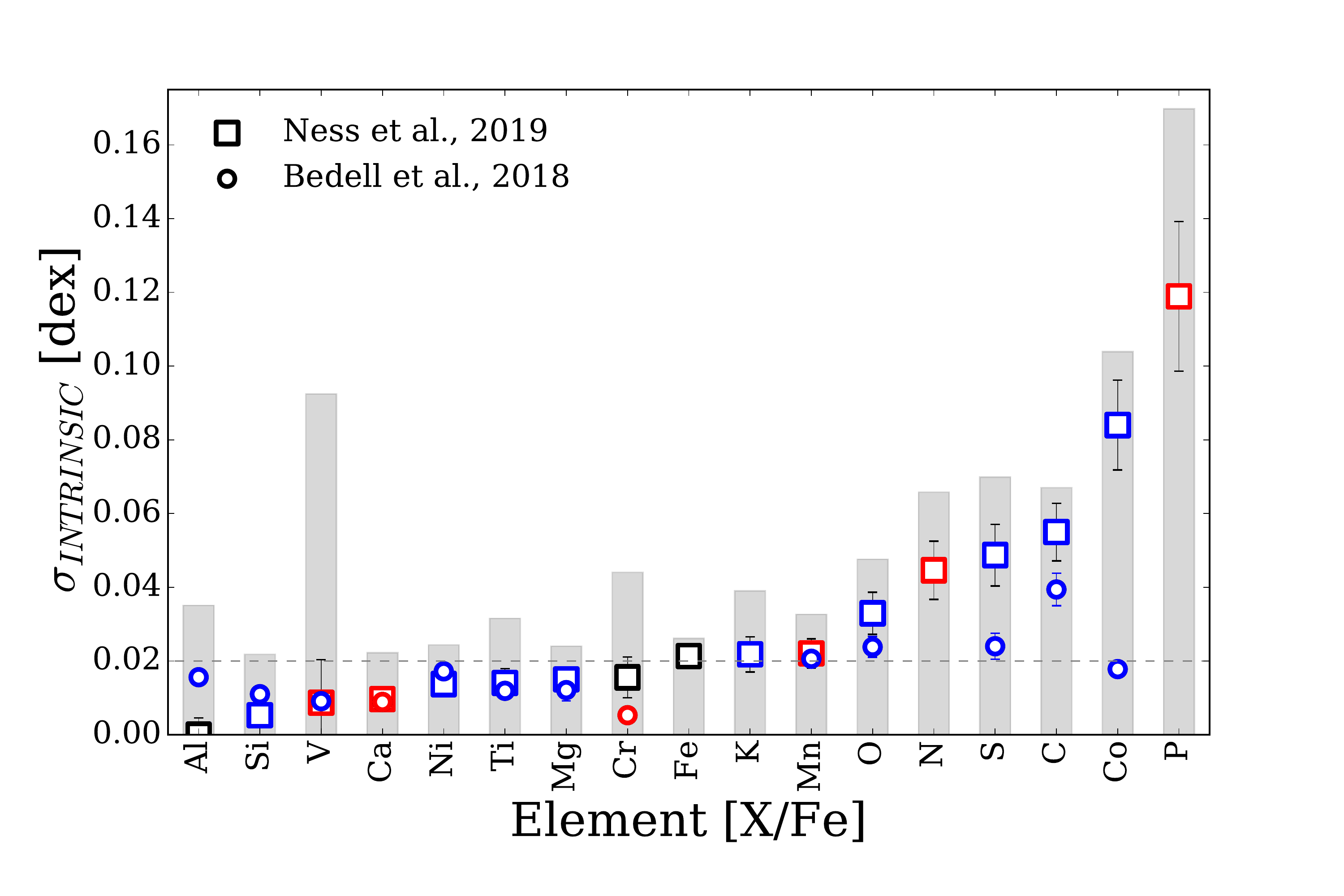} 
\caption{The intrinsic dispersion for stars at \feh\ =  0 $\pm$ 0.05 dex, measured for each element in Figure \ref{fig:ageabund2a} around the age-abundance relation fit by a 2nd-order polynomial model to the data, as shown by the square markers. The elements are arranged in order of increasing intrinsic dispersion. This intrinsic dispersion, $\sigma_{INTRINSIC}$,  is the dispersion around the age-abundance trend not explained by the measurement error, $\sigma_{MEASUREMENT \: ERROR}$. The grey filled bars show the total mean \textit{rms} dispersion for each element around the best-fit line ($\sigma_{TOTAL}$). Our results for the red clump stars are very similar to the intrinsic dispersion measurements for the solar-twin study of \citet{Bedell2018}, also shown for comparison, in the round markers. The markers are colored according to the sign of the gradient calculated from a linear fit to the data; black is consistent with zero, red is negative and blue is positive. The dashed line indicates the mean intrinsic dispersion measure for the elements, of $\approx$ 0.02 dex. }
\label{fig:ageabund2b}
\end{figure*}

For these set of stars shown in Figure \ref{fig:ageabund2a}, we calculate the intrinsic dispersion of each element, around the best-fitting line describing the age-abundance relation. This intrinsic dispersion is the amount of scatter that is not explained by the measurement error due to the abundance and age uncertainty. The total dispersion around the abundance-age relation is simply the \textit{rms} sum of the dispersion expected given the measurement error and the underlying intrinsic dispersion for that element around a given age: \\

$\sigma^2_{TOTAL}$ = $\sigma^2_{MEASUREMENT \: ERROR}$ + $\sigma^2_{INTRINSIC}$ \\

We therefore calculate the intrinsic dispersion quantity, $\sigma_{INTRINSIC}$ at a given age, for each element, by taking the \textit{rms} difference of the mean \textit{rms} dispersion around the line of best fit to the age-abundance trend and the mean measurement error.  The mean measurement error, which contributes to the scatter around the best fitting age-abundance line, is calculated as the \textit{rms} combination of the mean measurement errors on the abundances and the propagated age error that contributes to additional scatter around the best fit line to the age-abundance relation.  This (mean) quantity is shown in the top right hand corner of each sub-panel in Figure \ref{fig:ageabund2a}.  The contribution of the age-error to the uncertainty is determined by drawing a new age for every star from its Gaussian distribution, given its age and age error, and calculating the subsequent total \textit{rms} dispersion away from the already determined 2nd-order polynomial best fit through the data, for this new distribution of ages. This is repeated 100 times. The mean of the calculated dispersion around the best fit line from these draws is taken as the contribution of the age uncertainty to the dispersion accounted for by the age measurement error around the line of best fit. This value is typically $<$ 0.01 dex and decreases with a decreasing age-abundance correlation.  The intrinsic dispersion is then determined as the \textit{rms} difference of the 1$-\sigma$ dispersion measured around the line of best fit to the age-abundance relation (shown in the bottom right hand corner of each sub-panel) and the 1$-\sigma$ dispersion of the mean joint measurement errors from the age and abundance uncertainties.

Figure \ref{fig:ageabund2b} shows the measured intrinsic dispersion for each element (in square scatter points), ordered in increasing value of intrinsic dispersion as measured for our sample. A dashed line is drawn through the median intrinsic dispersion value, at  $\approx$ 0.02 dex. The x-axis lists all elements with respect to [X/Fe], expect for Fe which is \feh. The filled grey bars for each element show the  total 1-$\sigma$ dispersion around the best fit age-abundance polynomial. The color of the square markers indicates the overall gradient of the calculated intrinsic dispersion: blue represents a net positive age-abundance gradient, red represents a negative gradient and black represents a gradient consistent with zero, within the errors. For comparison, the measured intrinsic dispersion values from a set of $\approx$ 100 solar-twin stars (around \feh\ $\approx$ 0 dex) are shown in the circles \citep{Bedell2018}. The intrinsic dispersion values we measure from the red clump giants are remarkably similar to the main sequence solar-twin sample of stars from \citet{Bedell2018}, for many elements. There are some elements with marked differences, however, most notably Co and S. Similarly to the solar-twin study, these stars are restricted in their spatial extent to the \kepler\ field. Na and Rb are excluded from our analysis, as they are poorly measured (see Section 2). For Na we note the age-abundance trend is also counter to the expectation of other studies \citep[e.g.][]{Bedell2018} and the behaviour in our sample suggests this in fact would be consistent with being a noisy measurement of [N/Fe]. Therefore, we do not wish to draw conclusions regarding this element.

 Our abundance errors are important in the calculation of our intrinsic dispersion results, and we have taken care to ensure these are the most accurate estimates given our data. We find that the elements Al, Si, V, Ca, Ti, Na, Cr and Mg all have lower than mean intrinsic dispersion values. This is not necessarily surprising for Si, Ti, Mg and Ca: these are $\alpha$-elements and we have conditioned on the $\alpha$-element abundance of stars, selecting only those stars in the low-$\alpha$ sequence. Thus, selecting on a narrow range of these abundances renders them only fractionally additionally informative beyond being indicators of age. The elements K, Mn, O, N, S, C, Co and P all have intrinsic dispersion measures that are larger than the mean of the set of elements, up to the highest value measured of $\approx$ 0.12 dex for P. The intrinsic dispersion results are however very sensitive to the accuracy of our uncertainty. The element that is measured most imprecisely, P,  has the highest intrinsic dispersion. This may be a consequence of any inaccuracy in the uncertainty estimate which will propagate with a higher amplitude in the intrinsic dispersion calculation, of the \textit{rms} difference of the total (larger scatter) and the total measurement error. No comparative measurement is available for this element from the solar-twin study. We note however that P is produced in massive stars and chemical evolution models can not reproduce the distribution of this element in the Milky Way \citep{P2012}. We have no reason to expect our measurement uncertainties are not accurate. We have taken into account the signal to noise dependency of the error and the parameter space of our red clump stars across \teff, \logg\ and [Fe/H]. Our findings of small intrinsic dispersion values for the elements around the age-abundance relations are consistent with what has been already found in the solar-twin study of \citet{Bedell2018}, where the mean intrinsic dispersion across 66 element measurements is $\approx$ 0.02 dex (and ranges from 0 - 0.06 dex).  

\subsection{Constructing a model of ages from abundances to measure intrinsic dispersions across (age, [Fe/H])} 

We now construct an abundance model to infer ages and overall metallicity, \feh, for the 15,000 \apogee\ red clump stars in the low-$\alpha$ sequence, that span Galactic radii 4 -- 16 kpc. 
We construct this model similarly to \citet[see][]{Ness2015}, but instead of making a spectral model to generate stellar flux given abundances, we make a model to generate the abundance vector given age and \feh. We use the $\approx$ 600 low-$\alpha$ red clump stars with ages measured from the \kepler\ mission as our training set, for which we have \apokasc\ ages, with fractional age errors of $<$ 30 percent and SNR $>$ 150, as well as high fidelity \feh\ measurements precise to $\approx$ 0.03 dex. From this training set, we make an abundance model, using the 18 abundance features of [C, N, O, Mg, Al, Si, S, K, Ca, Ti, V, Mn, Ni, P, Cr, Co, Rb, [C/N]], with their associated uncertainties. For our model labels, we take both age and \feh.  We exclude Na due its anomalous behaviour although note that including it does not change the results. We leave Rb in our set of abundances as although it is poorly measured as indicated in Section 2, our inference is marginally improved (at the 5 percent level) with its inclusion.

Our model is then characterized by a coefficient vector $\set{\theta}_i$ for our $i$ abundances, that allows us to predict each abundance $X_{ni}$ for our $n$ training objects, for a given label vector, $\starlabelvec_n$, of age, $\tau$ and overall metallicity, \feh. We write our abundance model as a linear function of a vector $\starlabelvec_n$ built from the labels: 

\begin{eqnarray}
X_{ni} &=&
\set{\theta}_i^T \cdot \starlabelvec_n + \mbox{noise}
\label{eq:linearmodel}\quad
\end{eqnarray}

In the equation above, $\set{\theta}_i$ is the set of spectral model coefficients at each [X/Fe] for the $i$ abundances. The noise is an \textit{rms} combination of the associated uncertainty variance, $\sigma_{ni}^2$ of each of the abundance measurements (itself an \textit{rms} sum of the cross-validation and formal uncertainty from the optimizer we use in python, curve$\_$fit) and the intrinsic variance or scatter of the model of the fit at each abundance, $s_i^2$. Our quadratic model fits a set of coefficients at each feature, in this case, each abundance measurement, given our labels, in this case, ages and \feh. The training stage optimizes not only for the coefficients, but also a scatter term. This scatter term is equivalent to the (normally distributed) deviations from the approximate model for the true feature (abundance) variance around the best fitting (2nd-order that we chose here) polynomial function to the data. This scatter is used by the model at the inference stage, as an effective weighting of the features to infer the labels. This model assumes that the noise model is:


\begin{eqnarray}
\mbox{noise} = [s_i^2+ \sigma_{ni}^2]\,\xi_{ni}
\label{eq:linearmodel2}\quad
\end{eqnarray}

\noindent where each $\xi_{ni}$ is a Gaussian random number with zero mean and unit variance.

We use a polynomial model of our two labels of age and overall metallicity, $\tau$ and \feh, such that there are 6 coefficients at each abundance feature $\starlabelvec_n$ = [1, \feh, $\tau$, (\feh\ $\cdot$ $\tau$), (\feh)$^2$, ($\tau$)$^2$], where the first element ``1'' permits a linear offset in the fitting. For the \feh\ and age labels $\tau$, these values are scaled to a mean of 0 and a variance of 1 in order to keep the model stable and pivoting around a reasonable point in label space. 

\begin{figure*}[]
\centering
\includegraphics[scale=0.3]{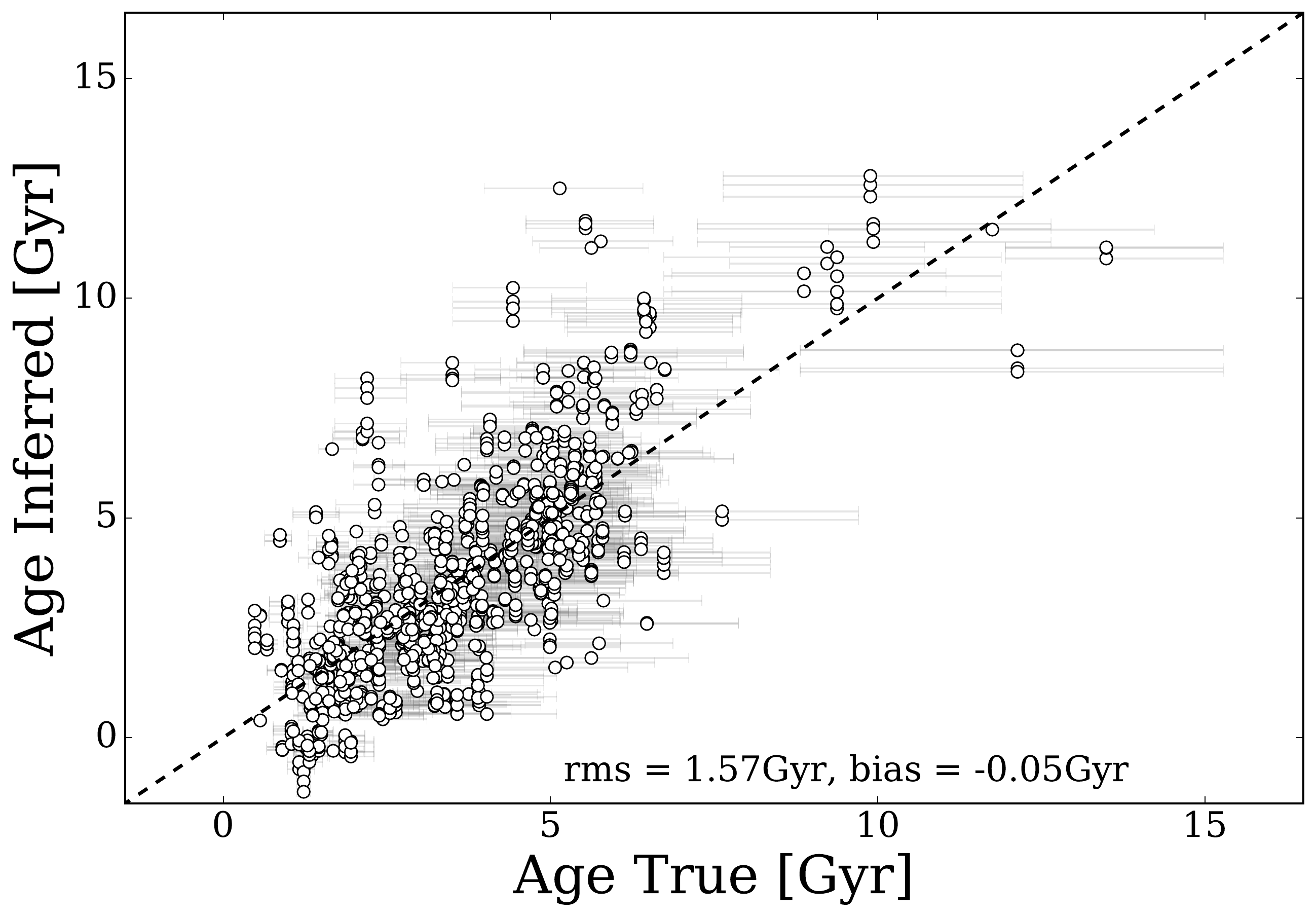} 
\includegraphics[scale=0.3]{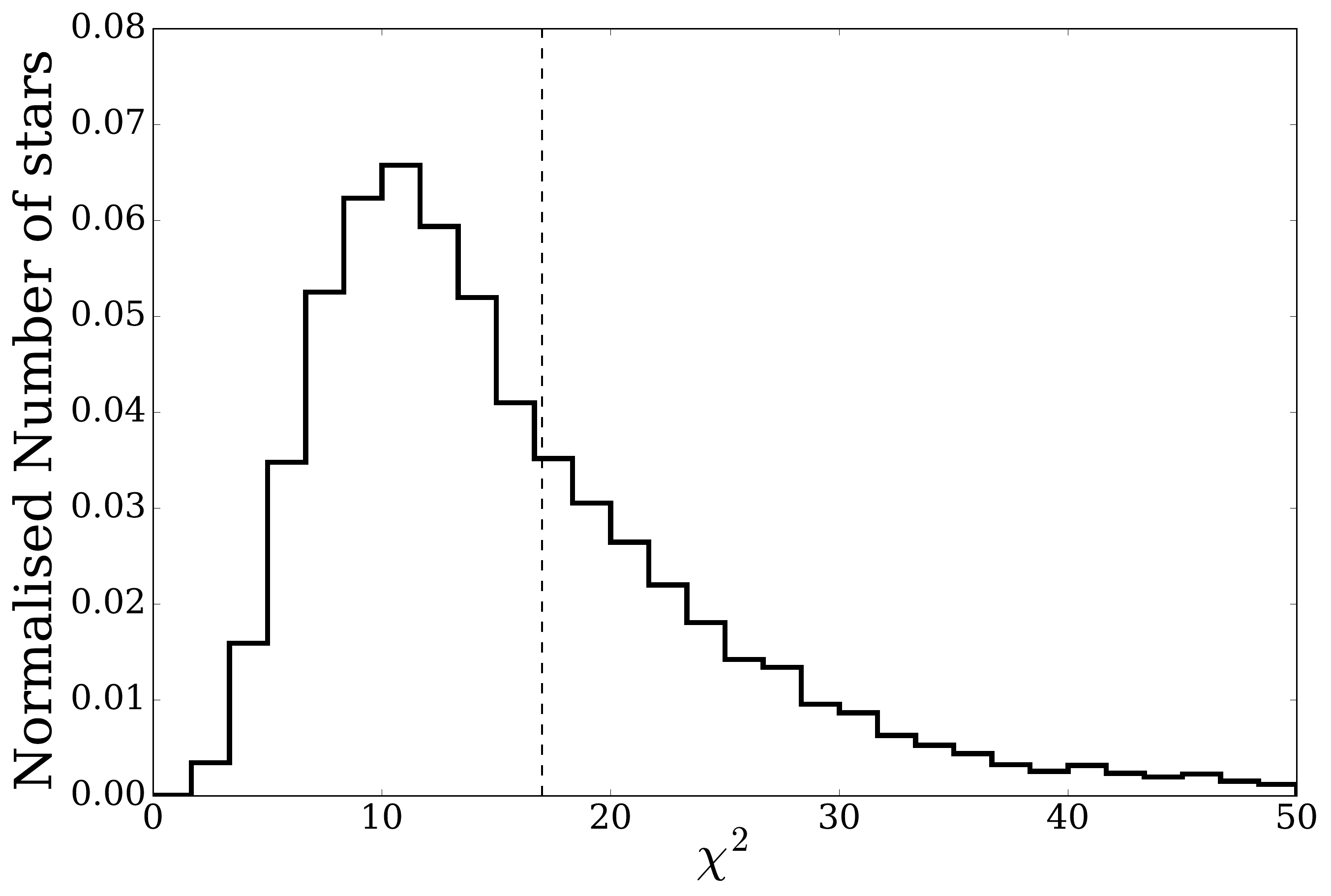} 
\caption{At left, the cross-validation results for our age inference using our quadratic model of the abundances given the ages and \feh, for our $\approx$ 600 stars in the training set (see Equation 1). We remove 5 percent of stars to create 20 models each with a different missing 5 percent to generate this Figure. Each of the 20 models is used to infer ages for the removed stars, which are shown. This training set of stars comprises stars with SNR $>$ 150 and asteroseismic age errors $<$ 30 percent. This model returns age precise to $\lesssim$ 1.6 Gyr, and metallicity, \feh\ to 0.07 dex precision (which is not shown). At right, the $\chi^2$ distribution of the model compared to the data for the $\approx$ 15,000 red clump stars in the low-$\alpha$ sequence for which we derive ages and \feh\ from our model. The dashed line indicates the $\approx$ expected peak of the $\chi^2$ distribution given the 18 abundance features that are used to infer the age and \feh\ labels. Our distribution peaks below this expectation which is likely a consequence of correlated features and errors, which are not accounted for in our simple $\chi^2$ metric. Nevertheless this is demonstrative that our model generates a good fit to the data for the red clump stars in the disk, with only a small tail of stars with high $\chi^2$ values where the model poorly fits the data.}
\label{fig:thecannona}
\end{figure*}

We test our model using cross-validation, training on 95 percent of the data and testing on 5 percent, which we perform 20 times. The result of this inference for age is shown in the left hand panel of Figure \ref{fig:thecannona}. The asteroseismic age is denoted as ``True Age" on the x-axis and our inferred age from our model is on the y-axis. We infer age to a precision of about $\lesssim$ 1.6 Gyr with almost no overall bias, although there is some apparent systematic effect at an asteroseismic (\apokasc) age of around 6 Gyrs, for which a range of ages are inferred,  up to $<$ 12 Gyr. We infer the metallicity, \feh, for the stars to an \textit{rms} precision of 0.07 dex with negligible bias. Our model is generative, which means we can generate the abundance vector given our ages and overall metallicity for each star, $\tau$ and \feh. The goodness of fit of our model compared to our data for the 15,000 stars in the low-$\alpha$ sequence for which we infer ages is shown in the right hand panel of Figure \ref{fig:thecannona}. We quantify our goodness of fit with a $\chi^2$ metric, where for each star: 

\begin{eqnarray}
\chi^2 = \sum_{i=1}^{18} \frac{ (X_{i_{model}} - X_{i_{data}})^2}{(s_i^2 + \sigma_{i}^2)}
\label{eq:chi2}\quad
\end{eqnarray}

\noindent for the i=1 to 18 abundance features.  The $\chi^2$ value for the full set of low-$\alpha$ red clump stars with --0.5 $<$ \feh\ $<$ 0.45 dex peaks at $\chi^2$ = 12 and 90 percent of our stars have $\chi^2$ values of $<$ 35, which indicates that our model is a good fit to the data for our sample, where we fit for 18 individual abundance features. Note the peak of the $\chi^2$ is lower than the nominal expectation from this many label variables (18), and we expect this to be a consequence of the element correlations. We exclude stars with a $\chi^2$ metric of $>$ 35 for our subsequent analysis. These more poorly fit stars represent only a small ($<$ 8 percent) fraction of stars for which the inferred age and \feh\ labels do not generate a good fit to the abundance data. This quality cut in $\chi^2$ ensures we only include ages where the generated model to the abundance data is a fairly good fit, at worst. We examined the $\chi^2$ fit of the model compared to the data in detail, and for a given age, this does not change significantly as a function of orbital property or Galactic location. This leads us to conclude that our local model of the age-abundance relation holds globally across the disk. If the low and high-$\alpha$ sequences were combined for such analysis this would not be apparent, as the high-$\alpha$ stars show a different abundance-age relation to the low $\alpha$-sequence of stars \citep[see][]{Bedell2018}. The relative fraction of high and low-$\alpha$ stars changes as a function of Galactic radii, which, combined, would look like a changing abundance-age relation across the disk which our local model would not be able to fit.

\begin{figure*}[]
\centering
\includegraphics[scale=0.5]{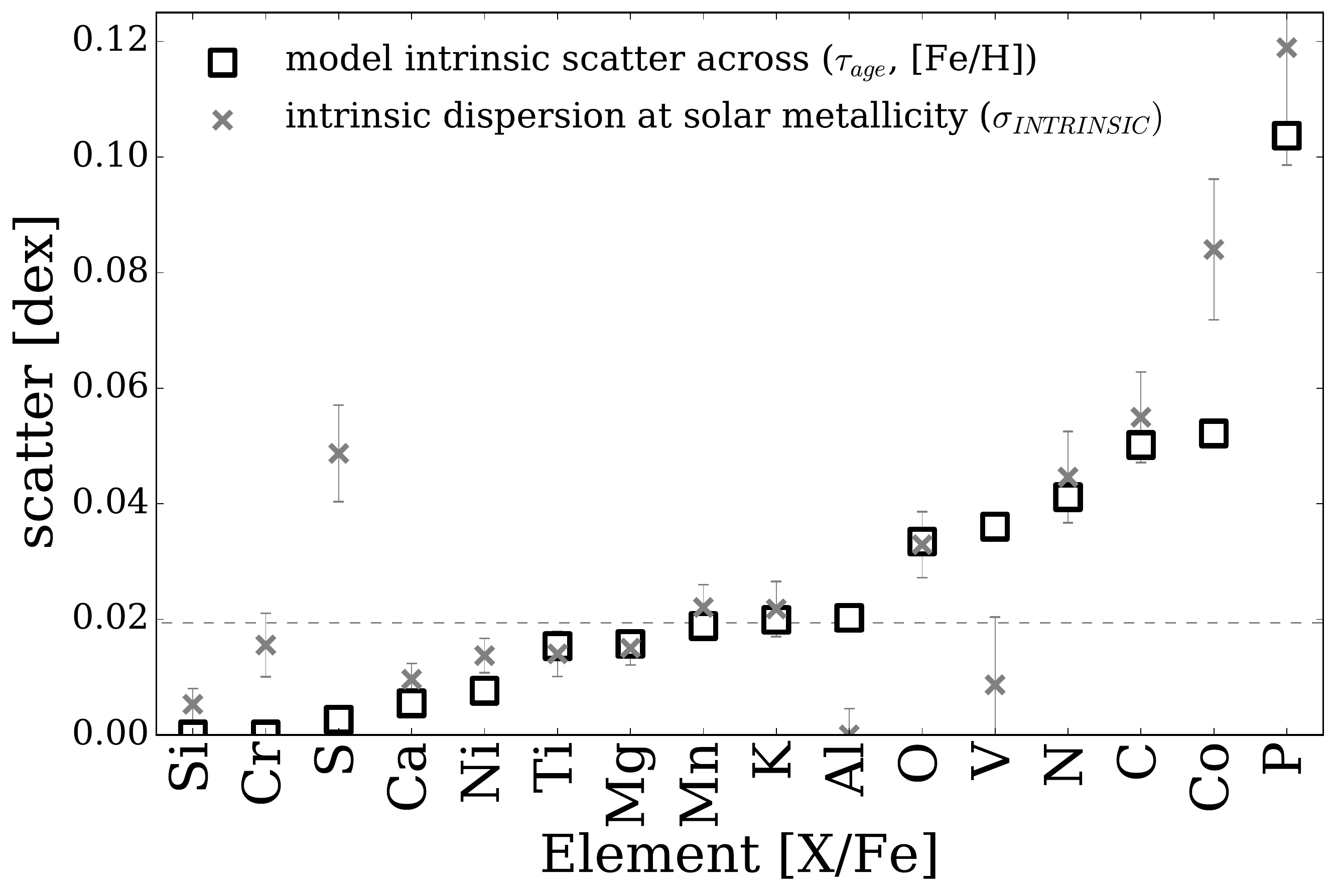} 
\caption{The model's scatter terms for each element, which parameterize how well the 2nd-order polynomial model fits the data for each element, given age and \feh\ labels. This model is constructed using $\approx$ 600 stars. The elements are arranged in order of increasing scatter and the magnitude of the scatter is indicated by the square markers. The scatter term represents the intrinsic dispersion of each element not accounted for by the error on the abundance label, when fitting for \feh\ and age simultaneously. The intrinsic dispersion of each abundance around the abundance-age relation at solar metallicity calculated from $\approx$ 100 stars is shown in grey crosses for comparison (from Figure \ref{fig:ageabund2b}). }
\label{fig:scatterthecannon}
\end{figure*}

Figure \ref{fig:scatterthecannon} shows the scatter, $s$, for each element from our quadratic model fit (see Equation 2), in order of increasing scatter (shown in black squares). The scatter term is effectively a global measure (across all \feh\ and age) of the overall intrinsic dispersion of each element; measuring how much additional variance around the model of abundances given ages and \feh, there is not accounted for by the (abundance) label errors. The intrinsic dispersion reported in  Figure \ref{fig:ageabund2b} by comparison is only for solar metallicity stars, \feh\ = 0 $\pm$ 0.05 dex. Here, we are generating the abundances using both age and \feh\ simultaneously. The scatter shown in Figure \ref{fig:scatterthecannon} is formally an upper limit of the intrinsic dispersion, given our 2nd-order polynomial model, as we do not take into account the contribution of the age errors in our model.  However, the contribution of the age uncertainties are small for the \apokasc\ sample. We ensure this by excluding stars with high age errors of $>$ 30 percent, and the average age error is 0.7 Gyr, which for the weak age-abundance relations does not propagate into a significant contribution to the total measurement uncertainty. We confirmed in our analysis in Section 3.1, that the age uncertainties have $<$ 0.01 dex  contribution to the total dispersion around the best fitting 2nd-order polynomial age-abundance fit to solar metallicity red clump stars. We compare the scatter from our model across age and \feh\ to the intrinsic dispersion we find for the solar metallicity stars in Section 3.1 in Figure \ref{fig:scatterthecannon}  (shown in grey crosses). In general these are identical or similar, with the exceptions of S, Al, V and Co.  

The goodness of fit of our generated model is not a function of the orbital or radial properties of our sample. That is, the model fits the stars near the Sun as well as the stars in the inner and outer disk.   Therefore, we can also conclude that our model is a good approximation for the age-abundance relations across the disk. Correspondingly, the scatter measurement for each abundance not only represents the intrinsic dispersion at a given (age, \feh) but also,  across the radial extent of the disk from 4 -- 16 kpc, which these stars span. This is demonstrative that moving beyond the solar neighbourhood (i.e. the solar-twin study of \citet{Bedell2018} and the analysis in Section 3.1 and Figure 4), the chemical evolution of the low-$\alpha$ disk has been largely homogeneous and consistent with what we see in the solar neighbourhood. This places strong constraints on mixing and gas accretion in the disk.

\subsection{The age-abundance trends for stars across the disk}

We have generated a model to infer age and \feh\ given abundances and subsequently determined ages for the red clump sample of stars which cover the parameter space of our training set. The high fidelity set of stars we can model well (with $\chi^2$ $<$ 35) corresponds to about 15,000 stars which are shown colored by age in the right hand panel of Figure \ref{fig:2panel_first}. Our cross-validation test using the training set described in Section 3.2 indicates that these inferred ages are precise to $\lesssim$ 1.6 Gyr (see Figure \ref{fig:thecannona}). From Figure \ref{fig:2panel_first} there is clear trend (even for the low-$\alpha$ stars only) with [Mg/Fe], with the abundance value increasing with age, which is indicative as to why $\alpha$-element enhancement has been a long-used age proxy \citep[e.g.][]{Haywood2013, Adi2013, VSA2018} (although more typically to distinguish ``young'' $\alpha$-rich and ``old'' $\alpha$-poor stars). However, at any given [Mg/Fe] value, there are a range of ages spanned, with age increasing as a function of \feh\ at a given [Mg/Fe]. At any given \feh, there is also a range of ages spanned, from the highest to lowest metallicity stars. 

\begin{figure*}[]
\centering
\includegraphics[scale=0.7]{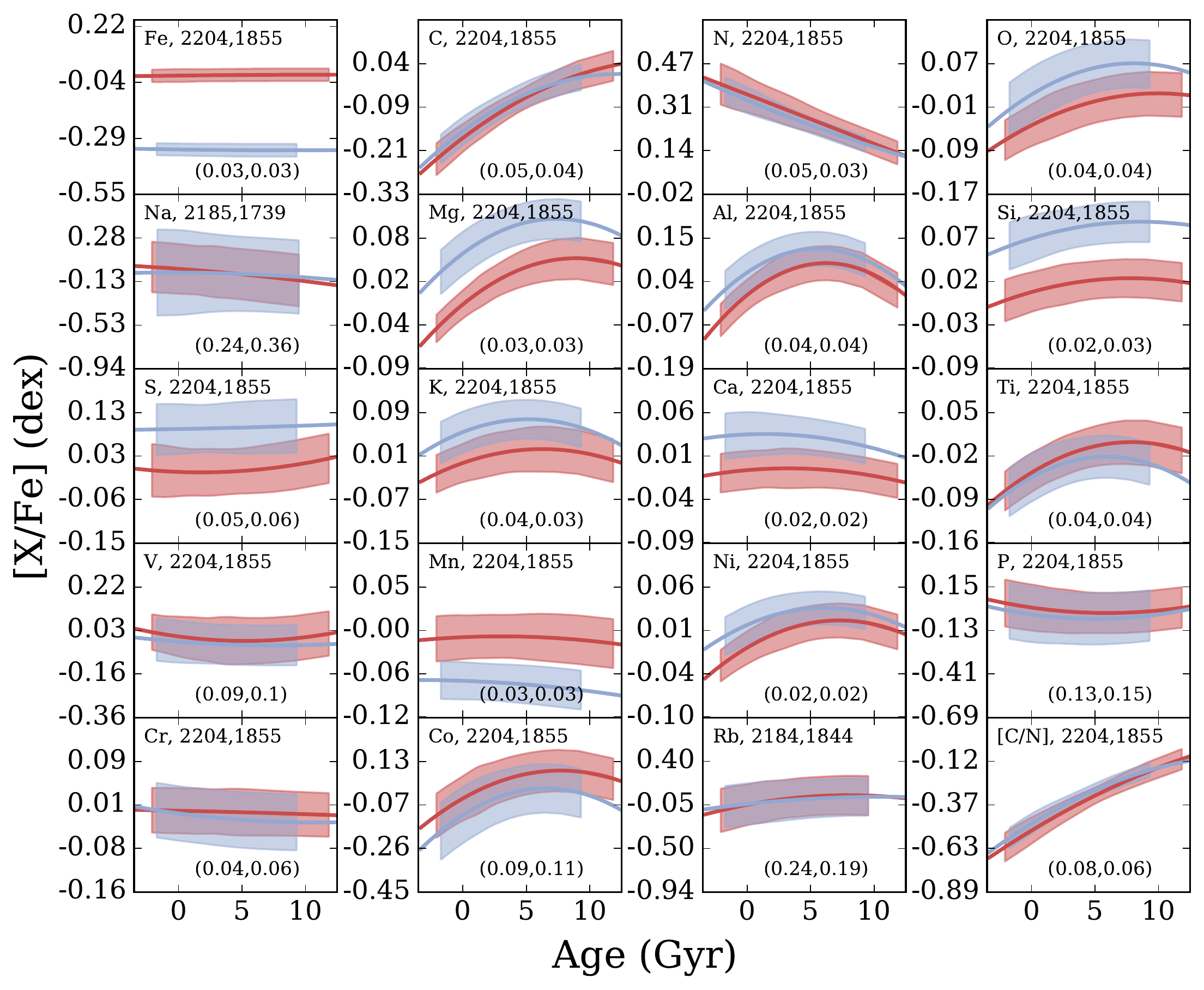} 
\caption{The best fit 2nd-order polynomial lines to the $\approx$ 1800 stars in our red clump sample with \feh = --0.35 $\pm$ 0.05 dex (in blue) and the $\approx$ 2200 stars with \feh = 0 $\pm$ 0.05 dex (in red), with the number of stars used to determine these lines indicated in each sub-panel. These stars shown within these narrow metallicity ranges are taken from the $\approx$ 15000 stars in the low-$\alpha$ sequence of the red clump stars with ages determined by modeling the relationship between the asteroseismic ages and abundances. The thicker regions around each line of best fit show the 1$-\sigma$ dispersion of the data. For most elements, the dispersion values for  metal-rich and metal-poor stars are comparable. However, the mean value of any element at a given age is dependent on the overall \feh, and for most elements, and the abundance-age slopes are typically steeper for the more metal poor stars.  Only stars with a $\chi^2$ $<$ 35 are shown. Above this $\chi^2$ threshold we deem the model to be a relatively poor fit to the data.  }
\label{fig:allrc}
\end{figure*}

We show the age versus abundance results for metal-rich compared to more metal-poor stars, in Figure \ref{fig:allrc}. In this Figure we show the best fit age-abundance trends (again using a 2nd-order polynomial model) for $\approx$ 1800 stars with \feh = -0.35 $\pm$ 0.05 dex and $\approx$ 2300 stars with \feh = 0 $\pm$ 0.05 dex. The 2nd-order polynomial fit lines to the age-abundance trends for element for the two metallicity bins are shown in red for the metal-rich stars (\feh = 0 $\pm$ 0.05) and in blue for the metal-poor stars (\feh = -0.35 $\pm$ 0.05). The running mean of the 1-$\sigma$ dispersion of the abundance measurements are shown in the thicker shaded regions either side of the best-fit lines, coloured red and blue respectively. It is clear from this large set of stars (more so compared to the analysis in Section 3.1 and Figure \ref{fig:ageabund2b}), that for many elements, the age-abundance trend is not described by a simple linear function. Most elements increase and flatten (e.g. C) or even turn over (e.g. Ni) at older ages. The age-abundance relationships also vary in amplitude and slope, as a function of \feh. This Figure also reveals that there are small differences in the dispersion values of [X/Fe] across age, as a function of \feh. 

\begin{figure*}[]
\centering
\includegraphics[scale=0.5]{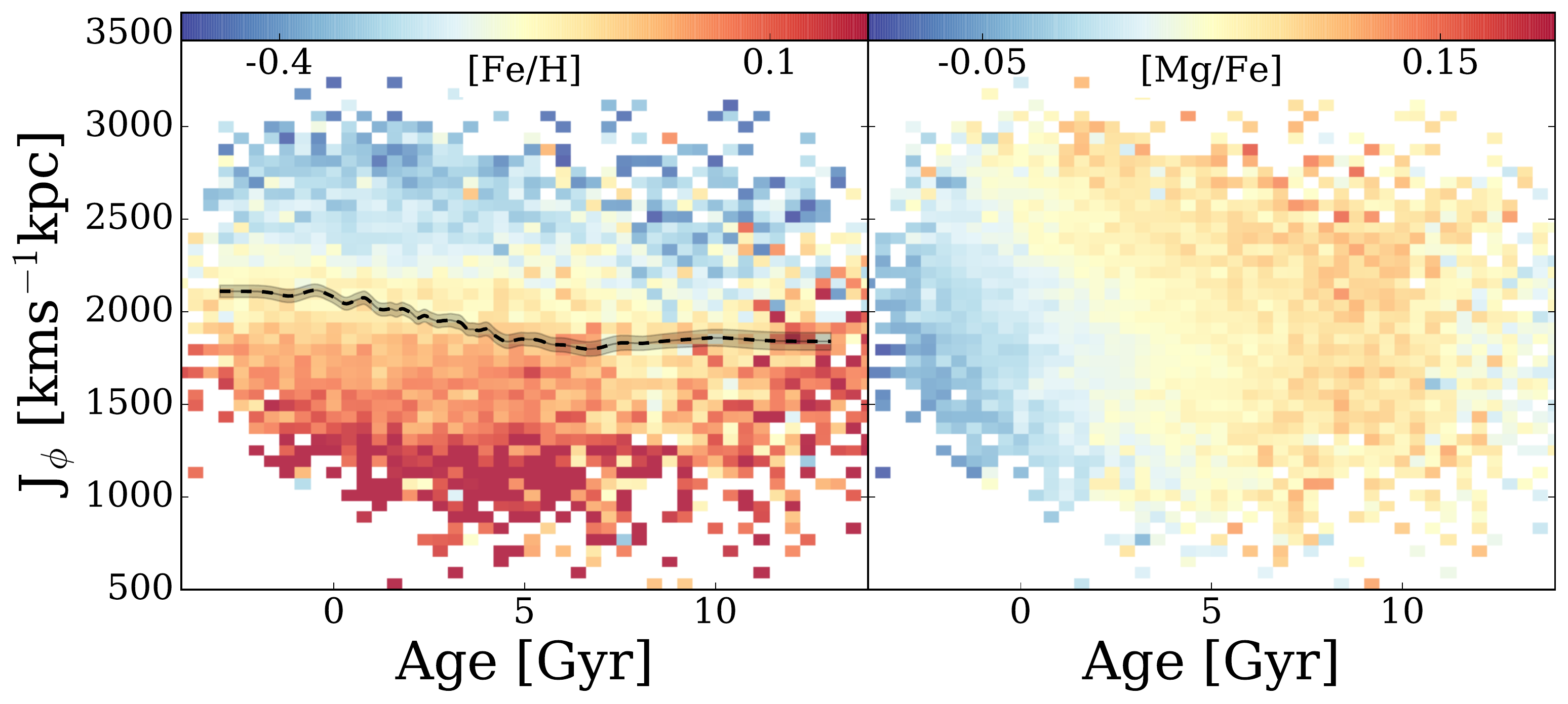} 
\caption{This Figure summarizes the abundance projections (\feh, $\alphafe$) into the age-$J_\phi$ plane, for the 15000 low-$\alpha$ red clump stars for which we have determined ages. At left, we show this plane coloured by \feh\ and at right, by [Mg/Fe], where the colour bar is shown at the top of each sub-panel. The dashed line in the left-hand panel shows the running mean of $J_\phi$ across age, which reveals the gradient across age in $J_\phi$: the mean age of the disk decreases with increasing Galactic radius. The left-hand panel shows that \feh\ is strongly correlated with $J_\phi$: the disk has a large overall metallicity gradient that decreases with increasing Galactic radius. At fixed $J_\phi$ there is also an \feh\ gradient across age: younger stars are more metal-rich than older stars at fixed $J_\phi$.  The right hand-panel shows that Mg-enhancement is an excellent age proxy: at fixed $J_\phi$ there is a strong gradient with [Mg/Fe] across age. At fixed age, there is also a weak trend in [Mg/Fe]: the [Mg/Fe] abundance of stars increases with increasing radius. }
\label{fig:gradients}
\end{figure*}

Taking [O/Fe] as an example from Figure \ref{fig:allrc}, this element shows metallicity dependent age-abundance trends. The more-metal-poor stars have a higher O enhancement and show a steeper rate of abundance decrease with age compared to the more metal-rich stars. By comparison the iron-peak elements V and Cr show basically no differences for the two metallicity bins. Of the $\alpha$-elements, O, Mg, Si, Ni and Ti show positive age-abundance relationships and Ca shows a negative age-abundance trend. Of all $\alpha$-elements, Ti is the only element where the metal-poor stars are less Ti-enhanced than the metal-rich stars and show a flatter age-abundance slope. The element Si shows the most separation of the $\alpha$-elements in the age-abundance relation between the two metallicity bins. This highlights that $\alpha$-elements have individual age-abundance behaviors as a function of both \feh\ and age, and themselves trace the detailed chemical evolution history of the low$-\alpha$ sequence of stars. Such comparative relationships are extraordinarily constraining for chemical evolution models and yield tables.  Figure \ref{fig:gradients} also highlights that for studies of the disk, if stars are not considered in bins of both age and \feh, then the net effect of the different trends with \feh\ and age will be to dilute the underlying relationships that carry the information about the detailed star formation history, which we can see here is clearly richly described in the data. We reiterate this is for the low-$\alpha$ sequence of stars only, the full distribution of stars increases the complexity of these relationships and proceeding by considering them separately is well motivated given their likely entirely separate origins \citep[e.g.][]{Gandhi2019, Ted2019a, Clarke2019}.

\section{Angular momentum gradients with abundances in the low-$\alpha$ disk}

We have investigated and shown the age-abundance trends for our set of red clump stars for which we have inferred ages precise to $\lesssim$ 1.6 Gyr. We now wish to investigate the relationship between age, abundance and stellar orbital properties.

\begin{figure*}[]
\centering
\includegraphics[scale=0.7]{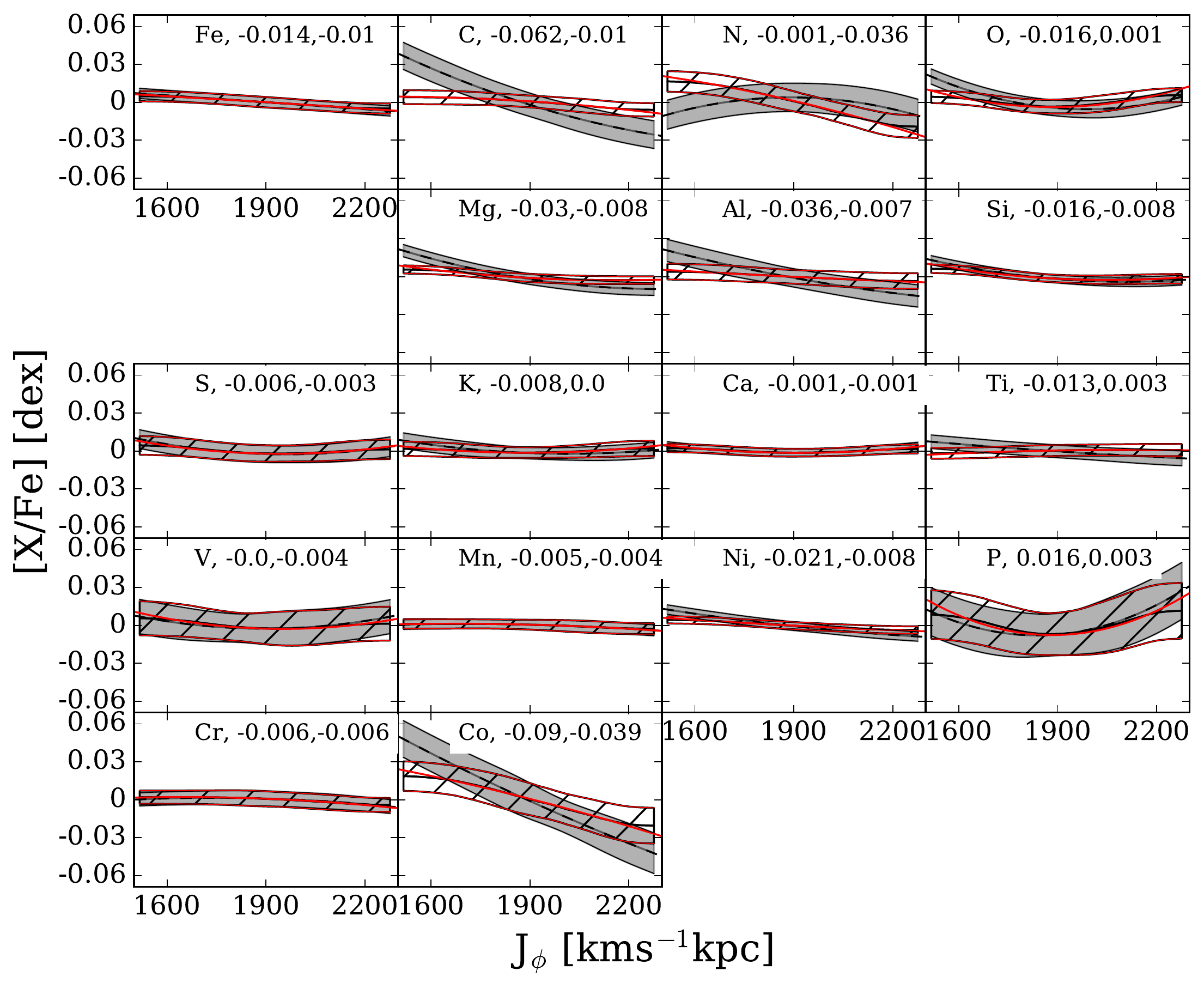} 
\caption{The $\approx$ 2000 red clump \apogee\ stars across the disk, with \feh\ = 0 $\pm$ 0.05 dex. The black dashed line shows the best fitting 2nd-order polynomial to the abundance-$J_\phi$ measurements for each element. The grey shaded region shows the 1-$\sigma$ dispersion of the data around this fit. The red line shows the best fitting 2nd-order polynomial to the difference between the age-abundance trend (shown in Figure \ref{fig:allrc}) and the value of each star's abundance (which we call the age-abundance residuals), and it's corresponding $J_\phi$ measurement, for each element. The red hatched region shows the 1-$\sigma$ dispersion of the data around this fit. The red hatched trends therefore represent the relationship between each element and $J_\phi$, with the age-abundance relation removed. For almost all elements, the removal of the age-abundance trends (almost entirely) flattens the correlation of the element with $J_\phi$, as shown in the best-fit to the age-abundance residual values in red. This demonstrates that the gradients seen between abundances and orbital radius in the disk are largely accounted for by the fact that elements indicate age. Elements with relatively high intrinsic dispersions, such as C and Co, however, show gradients in the residual line of best fit across $J_\phi$ that are not erased by accounting for the age-abundance relation, but likely link to differences in chemical evolution across the disk for stars of the same age. The trends shown above look very similar for the lower metallicity selection of stars (\feh = --0.35 $\pm$ 0.05 dex), shown in Figure \ref{fig:allrc}.}
\label{fig:allactions}
\end{figure*}

There is a known age gradient in the disk and in disks of spiral Galaxies \citep[e.g.][]{Ness2016, Gadotti2019}. Furthermore, there is a gradient in \feh\ that exists at fixed age \citep{Ness2016}.  There is also an established correlation between stellar $\alpha$-element abundances and age \citep[e.g.][]{Haywood2013,Adi2013}. In this work we show that most elements in fact correlate with age. From Figure \ref{fig:allrc}, we expect that any gradients that we see in element abundance enhancements across Galactic radius will in large part be due to the trends between abundances and age at a given \feh. 

We now have access to not only the stellar radius of our stars, which is their present-day orbital position, but from \textit{Gaia} proper motion measurements, their angular momentum, $J_\phi$, which enables are more accurate description of a star's average orbital radius in the disk. We use the angular momentum to describe this average stellar location, where small $J_\phi$ corresponds to the inner Galaxy, and large $J_\phi$, to the outer Galaxy. Our red clump stars span a Galactic radius of about 4-16 kpc and correspondingly, an $J_\phi$ range of about 500 - 3500 kms$^{-1}$kpc. Figure \ref{fig:gradients} shows the $J_\phi$-age distribution of the low-$\alpha$ stars coloured by \feh, at left, and [Mg/Fe], at right. We show [Mg/Fe] as this is representative of an $\alpha$-element, which is a well studied and mapped element across the disk, that is also been used as an age-proxy. The dashed line in the left-hand panel shows the running mean of $J_\phi$ as a function of age, which shows the age gradient across the disk, with the mean stellar age decreasing with increasing distance from the Galactic center. \footnote{This age gradient is fairly shallow, but note we are examining only the low-$\alpha$ stars. The age gradient is far steeper considering both low and high-$\alpha$ stars simultaneously, as the high-$\alpha$ stars are on average older and concentrated to the inner Galaxy where as the low-$\alpha$ stars are on average much younger and concentrated to the outer Galaxy.} The grey-shaded area around the line is the sampling error, representing the uncertainty of this mean $J_\phi$ measurement across age. The relationship between the abundance plane, $J_\phi$, and age is clearly demonstrated in this Figure. There is a strong trend between $J_\phi$ and \feh: this is the metallicity gradient in the disk with radius, that has been previously measured from Cephied tracers \citep[e.g.][]{Luck2011} and known to be around -0.06 dexkpc$^{-1}$, with the \feh\ increasing from the outer to inner disk (from larger to smaller $J_\phi$). The total mean change in \feh across $J_\phi$ = 1000 -- 3000 kms$^{-1}$kpc is $\approx$ 0.5 dex. This is consistent with the circular velocity of 220 kms$^{-1}$ \citep[e.g.][]{Paul2017}, where by the $J_\phi$ gradient for a star on a circular orbit will be $\approx$ -0.6 dexkpc$^{-1}$ / (220 kms$^{-1}$) = -0.27 dex/(1000 kms$^{-1}$kpc).

There is a very slight gradient with age in that at fixed $J_\phi$, higher metallicity stars are younger than more metal-poor stars. For the abundance enhancement [Mg/Fe], there is also a gradient seen between $J_\phi$ and [Mg/Fe] as a function of age: at fixed $J_\phi$ the [Mg/Fe] increases as stellar age increases. There is also a slight gradient seen across [Mg/Fe] with $J_\phi$: at fixed age, [Mg/Fe] decreases with decreasing $J_\phi$. 

We next examine the full set of element abundance enhancement-$J_\phi$ relationships for the red clump stars across the disk. Figure \ref{fig:allactions} shows the $\approx$ 2000 solar metallicity, \feh\ = 0 $\pm$ 0.05 dex stars, from our sample. For these stars, we show the 2nd-order polynomial line of best-fit to the abundance-$J_\phi$ measurements (in the black dashed lines). We also show the best-fit 2nd-order polynomial line (in the filled red lines in Figure \ref{fig:allactions}) to the difference between the age-abundance trend (shown in Figure \ref{fig:allrc}) and the value of each star's abundance (which we call the age-abundance residuals), and it's corresponding $J_\phi$ measurement, for each element. The grey shaded and red hatched regions around these lines are the 1-$\sigma$ dispersion of the measurements (for the abundance measurements and the residuals away from the age-abundance fit, respectively). The elements which we exclude as being unreliable are removed from this Figure, which is otherwise in the same order as the previous Figure \ref{fig:allrc}.  The magnitude of the abundance change over the range of $J_\phi$ = 1600 -- 2200 kms$^{-1}$kpc is shown in the top of each sub-panel for the abundance data and the residual data, respectively.  We expect and see that elements most correlated with age to show the largest gradients in their abundance-$J_{\phi}$ trends (i.e. C and N). We see  that for elements with small intrinsic dispersion values in particular, removing the age-abundance trend, as shown by plotting the best-fit line to the residuals, results in a flat abundance residual-$J_\phi$ trend compared to the abundance-$J_\phi$ trend.  We note that two of the elements with the highest intrinsic dispersion measurements (and scatters), Co and N,  show significant gradients in the residual-$J_\phi$ best-fit line. These gradients must not be a consequence of any age or \feh\ correlation with $J_\phi$.  Note that the elements Co and Fe are from the same nucleosynthetic, family. Yet, from this Figure and comparing the very different gradients we see for Fe and Co in the residual best fit lines across $J_\phi$, we can conclude that the element Co is clearly tracing some chemical evolution that Fe is not. This exercise highlights inter-element family uniqueness that can be exploited to isolate enrichment pathways, as explored within the $\alpha$-family of elements in \citet{Blancato2019}. 

\section{The Orbital Separation of Stars Using Abundances at Fixed \feh}

We have seen that the abundance architecture of stars is linked to their age and $J_\phi$. We now seek to examine if the detailed abundances of stars indicates their orbital properties. We proceed by examining the mean orbital actions of stars that are grouped by their abundance similarity, at a fixed \feh.  We take mono-\feh\ bins of our sample of disk stars. We take these bins across the metallicity range of -0.4 $\leq$ \feh\ $\leq$ 0.25 dex (where there are a sufficient number of stars for this test). The width of each of the \feh\ bins is approximately the error in [Fe/H]: $\sigma_{[Fe/H]}$ = 0.035 + 0.05 $\times$ $|$[Fe/H]$|$.  The mean radii for these metallicity selections spans from R$_{GAL}$ = 8.6 to 10.3 kpc (with the mean radii increasing from the more metal-rich to more metal-poor stars).  Here, we also condition on (i) vertical action $J_z$ $<$ 10 kms$^{-1}$kpc\footnote{This corresponds to a maximum vertical excursion z$_{max}$ $\sim$ 0.7kpc at \rgal\ = 8kpc and z$_{max}$ $\sim$ 0.78kpc at \rgal\ = 10kpc}, thus excluding stars that make excursions away from the plane of the disk and (ii) age, examining both only the youngest stars ($<$ 1 Gyr) and then comparing this to intermediate age stars ($>$ 5 Gyr).

\begin{figure*}[]
\centering
\includegraphics[scale=0.23]{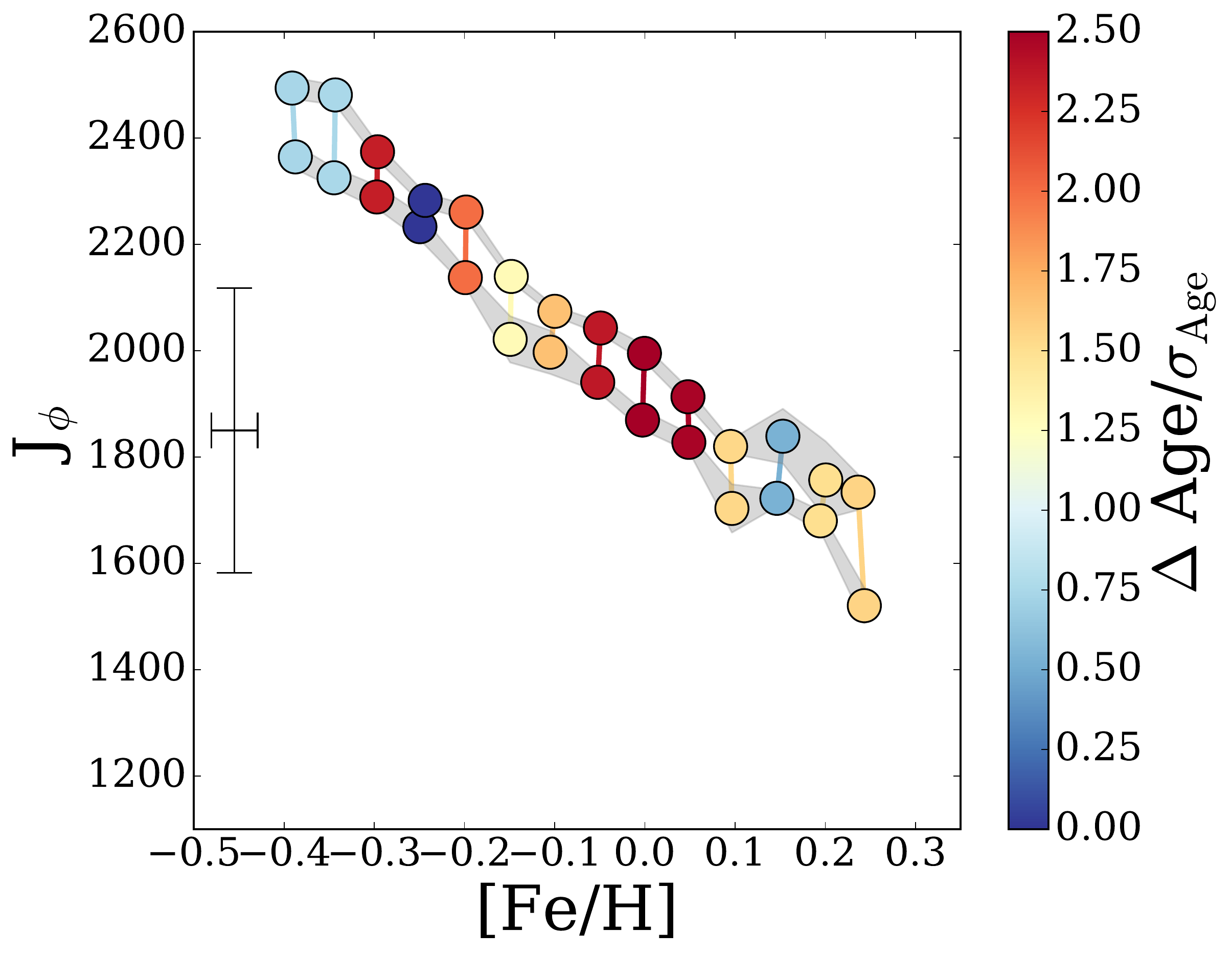} 
\includegraphics[scale=0.23]{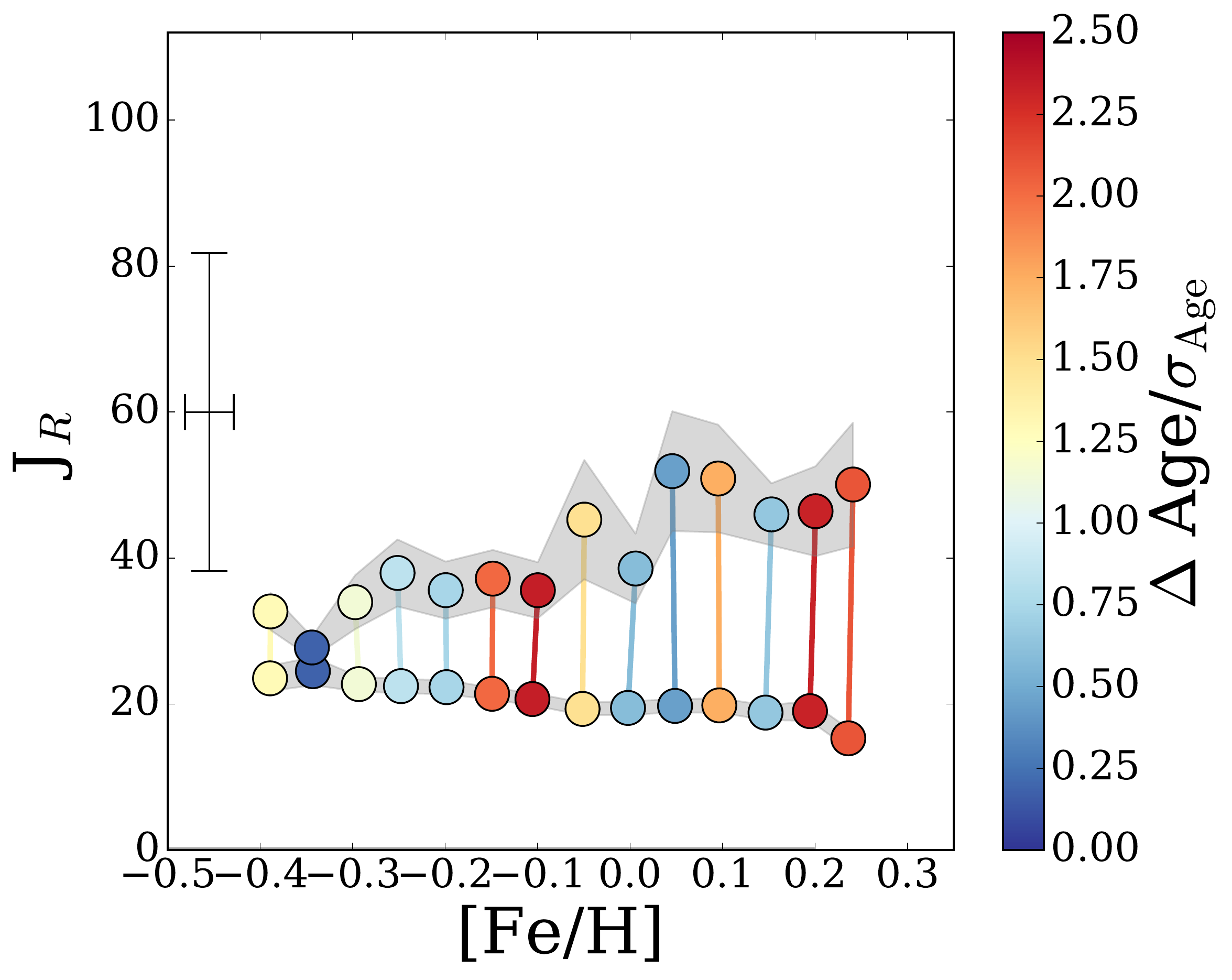} 
\includegraphics[scale=0.23]{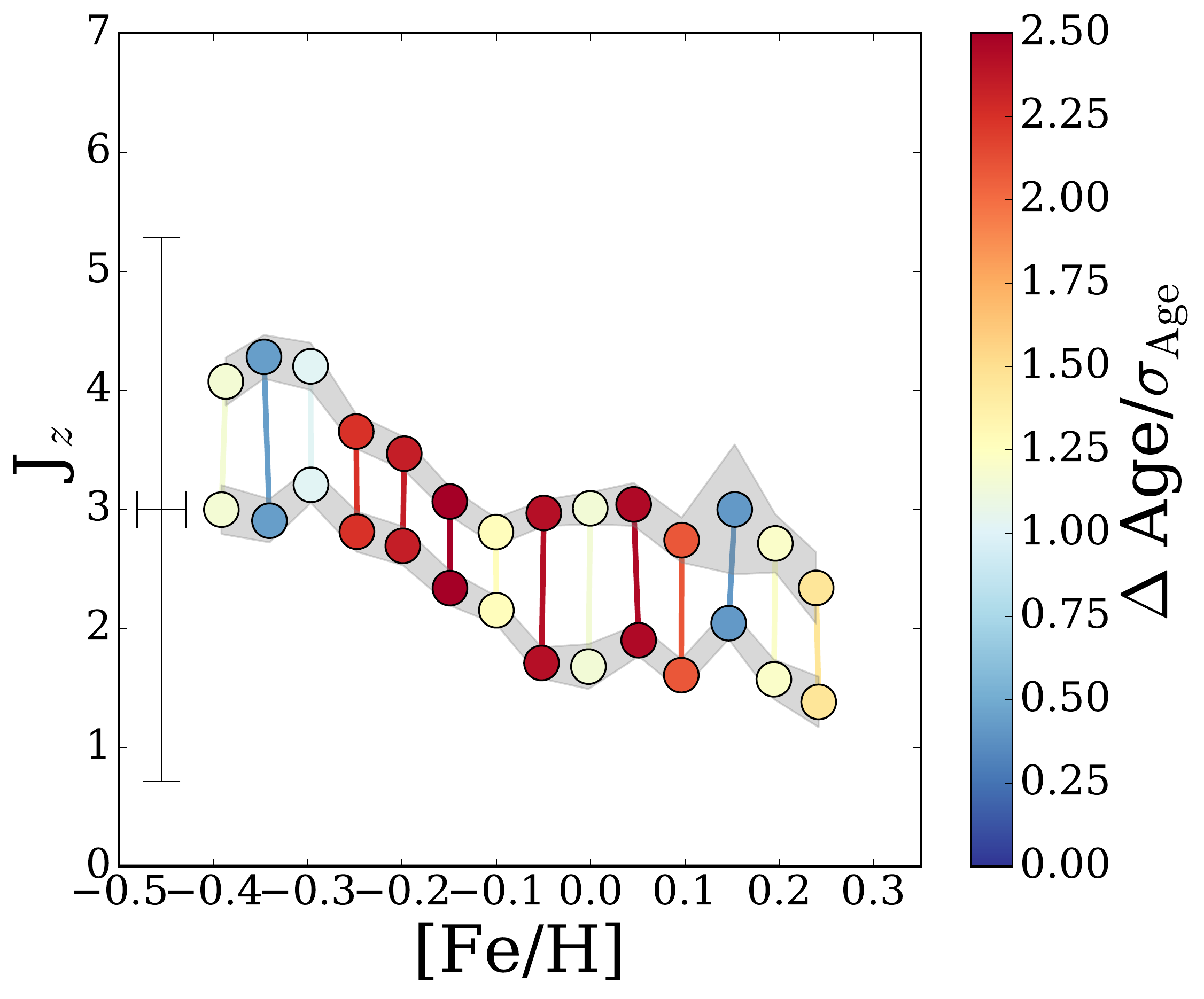} 
\caption{The mean actions ($J_\phi$, $J_R$, $J_z$), for low-$\alpha$ disk stars divided into two groups based on their element abundance similarity, [X/Fe], at fixed \feh. The division of the stars at each \feh\ into two groups is done using the k-means algorithm where each grouping is independent across \feh\ and each action. Each pair at fixed \feh\ is joined with a line. These groups are comprised of $\approx$ 6500 stars in total, with ages $<$ 1 Gyr and span an \feh\ range  --0.5 dex $<$ [Fe/H] $<$ 0.45 dex. The stars have also been selected to be confined to the disk, with $J_z$ $<$ 10 kms$^{-1}$ kpc. The circles, that show the mean action values for the two groups at each \feh, and the line that join them, are coloured by the age difference between the pair, normalised by the 1-$\sigma$ standard deviation of the age range of the stars in this selection.  The errors on each mean action measurement are shown in the thick grey line joining the open circles. This figure shows that populations of stars with different abundances, [X/Fe], at a fixed \feh,  separate out into groups of different mean ages, with different mean actions. The error-bar at left shows the average 1-$\sigma$ standard deviation of the individual action measurements of the stars that comprise the groups, around their mean.}
\label{fig:proof}
\end{figure*}

\begin{figure*}[]
\centering
\includegraphics[scale=0.23]{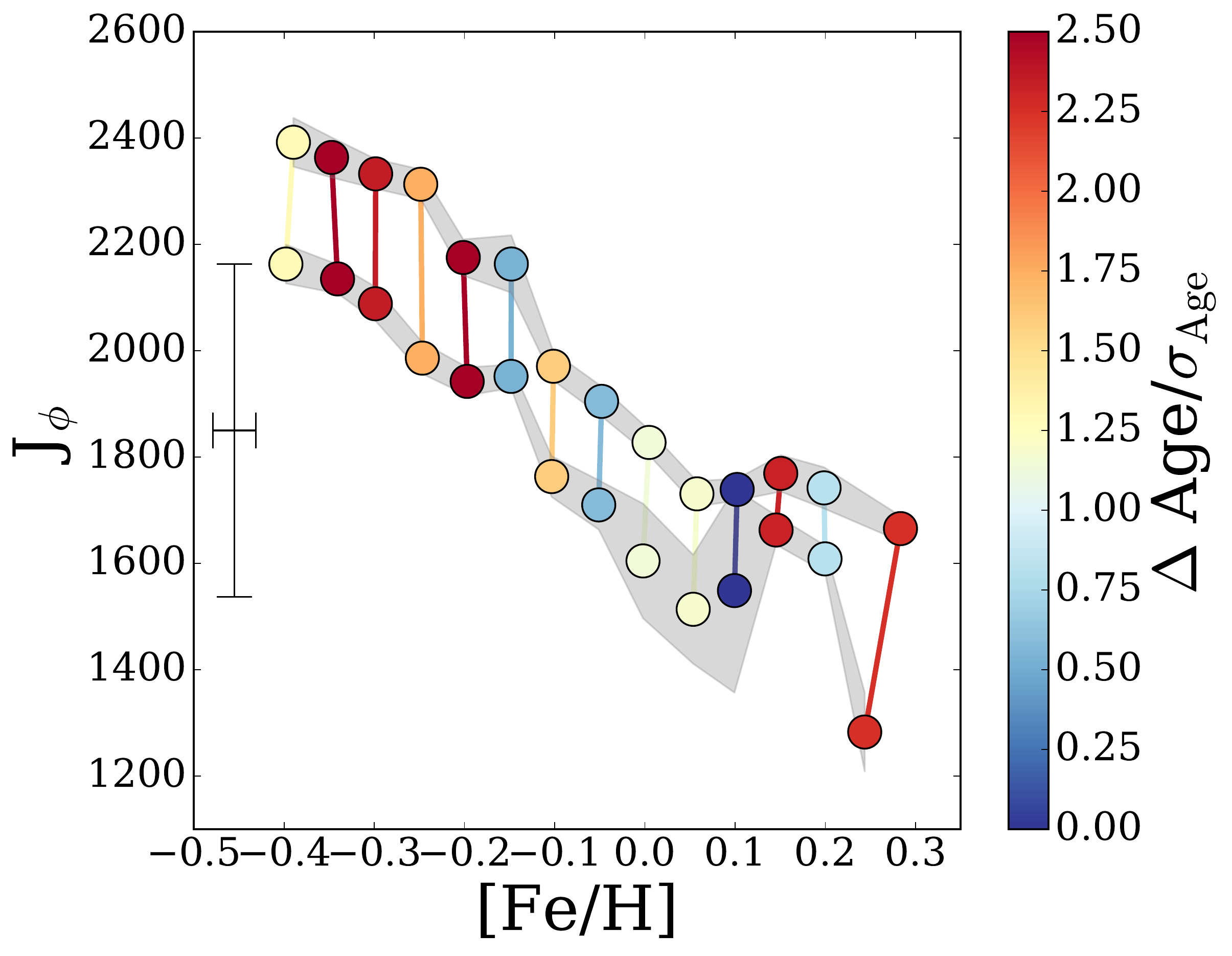} 
\includegraphics[scale=0.23]{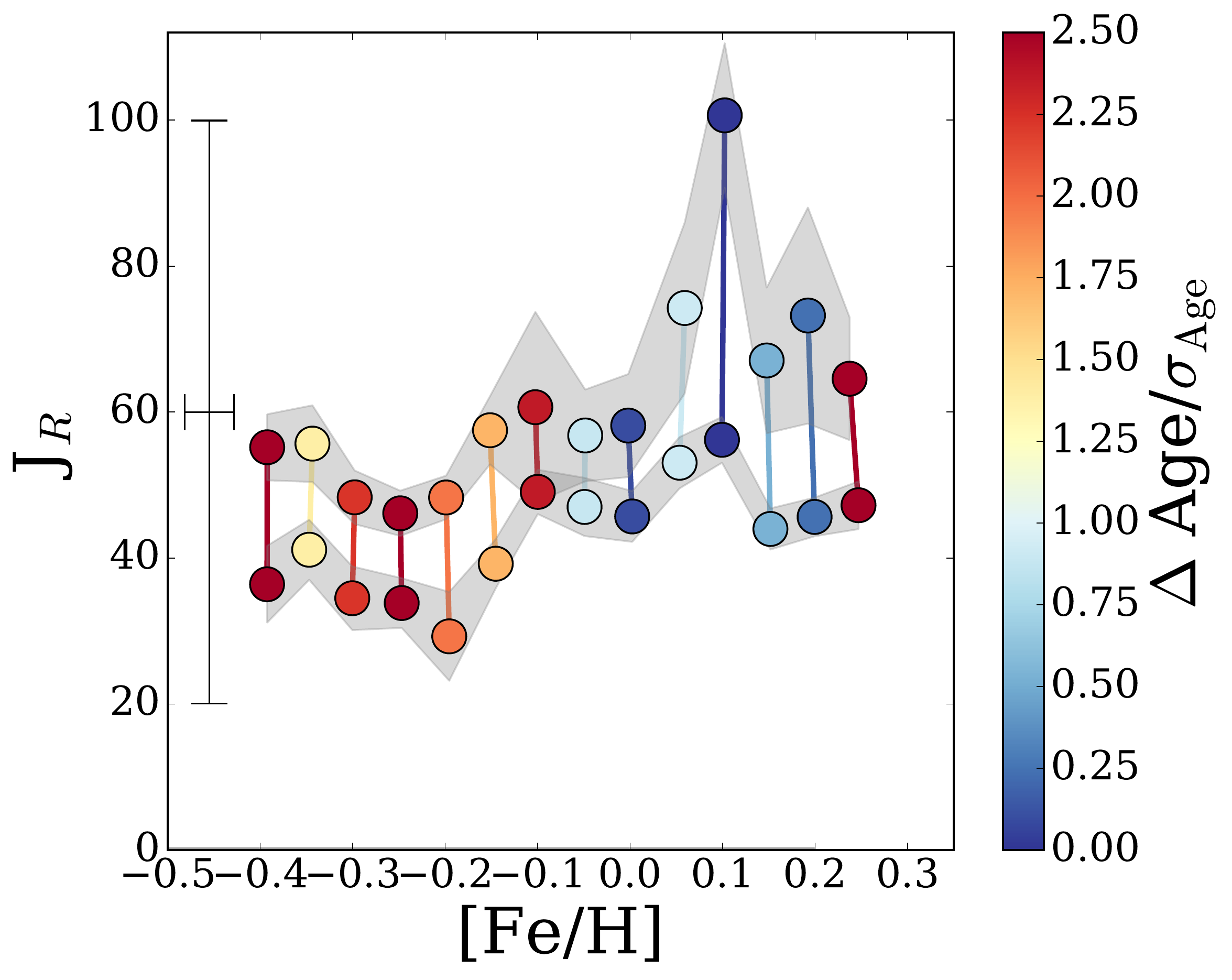} 
\includegraphics[scale=0.23]{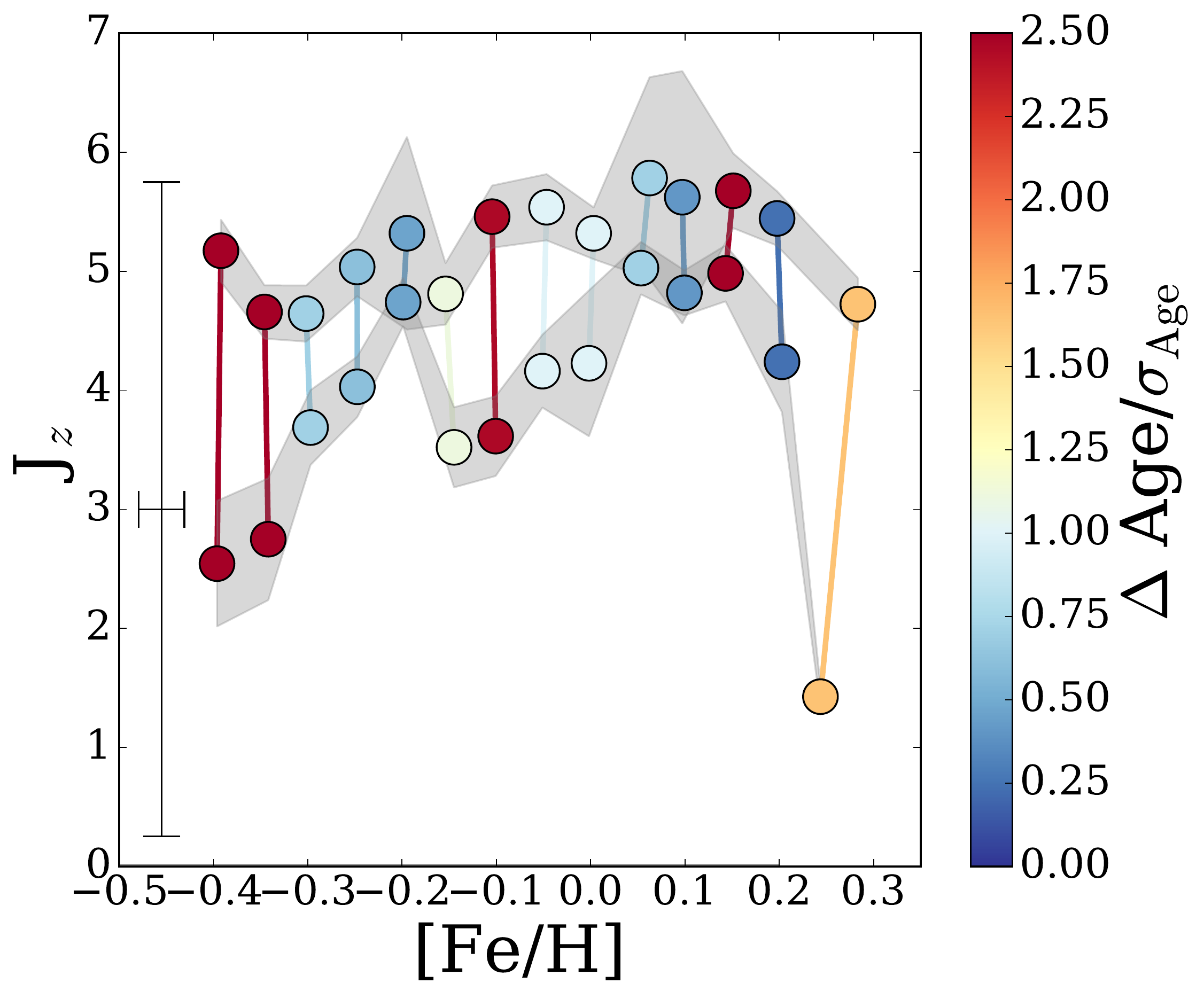} 
\caption{This is the same as Figure \ref{fig:proof} only taking the  $\approx$ 3500 intermediate-to-old stars with ages $>$ 5 Gyr, across the \feh\ range  --0.5 dex $<$ [Fe/H] $<$ 0.45 dex. Again, the stars have also been selected to be confined to the disk, with $J_z$ $<$ 10 kms$^{-1}$ kpc. As per Figure \ref{fig:proof}, this shows that populations of stars defined using their [X/Fe] abundances at a fixed [Fe/H] separate out into groups of different mean ages with different mean actions. Compared to Figure \ref{fig:proof}, older stars show greater separation in $J_\phi$, compared to younger stars, presumably as a consequence of having more  time to migrate from their birth origin \citep{Frankel2018}.  On average, comparing to Figure \ref{fig:proof}, both young and old stars shown similar average separation in $J_R$ and $J_z$. However, typically there is larger seemingly random variation in the action difference between the pairs for the older stars shown here compared to stars $<$ 1 Gyr.  }
\label{fig:proof2}
\end{figure*}

For the stars assigned within each metallicity bin, we use the k-means clustering algorithm and the [X/Fe] {information}, similarity to \citet{Hogg2016}, but here to separate the stars into \textit{two} groups at each fixed \feh, which we denote to be N=1 and N=2. Each grouping is independent, across \feh\ and across each action at fixed \feh. We allow each element to have a weighting coefficient and we optimize the coefficients for each action at each \feh, to maximize the separation between N=1 and N=2 for each orbital action (note that such a coefficient is not strictly necessary, but the otherwise equal weighting of the scaled abundances is arbitrary). These two groups, N=1 and N=2, represent populations of nearest neighbor abundance similarity and are determined from the [X/Fe] of the stars (not including, but rather at fixed \feh), for the low-$\alpha$ sequence (given our additional age and vertical action cut described). The N=1 and N=2 populations therefore represent groups of stars most chemically similar to one another:  stars within cluster N=1 and within cluster N=2 are more similar to each other than stars between the N=1 and N=2 clusters. We assert two clusters as we are restricted by our number of stars, but this exercise could be generalized to examining the orbital separation of an arbitrary number of groups. 

The mean actions, $J_\phi$, $J_R$ and $J_z$ for each of the two clustered assignments of stars into the N=1 and N=2 groups at each fixed \feh\ are shown in Figures \ref{fig:proof} and \ref{fig:proof2}, for each mono-\feh\ bin, for the stars with ages $<$ 1 Gyr and ages $>$ 5 Gyr, respectively. The x-axis for these Figures shows the mean \feh\ of the groups - where most groups have near identical mean \feh\ values, typically to the second decimal place, and the y-axis shows the mean action, respectively, for each action. The mean action values for each pair of stars within groups N=1 and N=2 are coloured by the age difference of the pair, normalized by the standard deviation of the age range of the stars (we do the normalization so as to use a consistent scale for our young $<$ 1 Gyr selection and older $>$ 5 Gyr selection of stars). 

Note that the members in each pair N=1 and N=2 are different for each action plot (with corresponding different mean ages and age differences). This is because for each mono-[Fe/H] bin for each action, we independently optimize the weighting of each element to maximize the action separation. The purpose of this is to test if abundances can indicate age and orbital properties for each component of the orbit (as opposed to testing if groups separated in say $J_\phi$ are discrete in their other orbital actions, which under radial migration we expect to not be the case). Note that we are also not trying to examine groups of stars of a common cluster birth origin, to test how their orbits may behave. We are not in the regime where we can reconstruct common birth sites of stars. We instead can examine if there is structure in the abundance-orbital plane, and we see that this structure links to stellar age. Without using the age measurements for the stars themselves, we can see that the vector of abundances reveals the structure in the dynamical configuration of the disk.

We find that which abundances are the most important to maximally separate each action (at each \feh)  changes, for each action. This must link to the initial radial, vertical and azimuthal properties of the gas disk from which the stars were formed, and the timescales of its enrichment and assembly. Over time, this link is confounded by how the stars have subsequently moved. 

Each pair of groups of stars, N=1 and N=2, are joined with a solid line, that is coloured by the normalised age difference of the pair.  The 1-$\sigma$ standard error on the mean action measurement of the groups is shown in the grey line joining the markers across all of N=1 and across all of N=2. The typical 1$-\sigma$ standard deviations around the mean action measurements are indicated at the far left of each sub-panel. 

In both Figures \ref{fig:proof} and \ref{fig:proof2}, the two groups N=1 and N=2, based on abundance similarity, have different mean ages for each pair (with a couple of exceptions). This is not surprising since abundances indicate age, as seen in Section 3. Groups based on similarity of abundances are effectively groups in age. The groups N=1 and N=2 in Figures \ref{fig:proof} and \ref{fig:proof2} also show significant differences in their mean actions.  Notably, the separation in $J_\phi$ between the two groups N=1 and N=2 is larger for the older stars $>$ 5 Gyr in Figure \ref{fig:proof2}, compared to the younger stars $<$ 1 Gyr in Figure \ref{fig:proof}. This is indicative of stars moving from their birth locations over time. The stars with ages $<$ 1 Gyr have simply had less time to migrate across the disk. Presumably, this separation of the groups N=1 and N=2 is a qualitative measure of the migration of stars over time. 

The $J_R$ result for young stars with ages $<$ 1 Gyr in Figure \ref{fig:proof} shows an increasing action separation between pairs of N=1 and N=2 groups for metal-rich compared to metal-poor stars. This trend is not seen for stars with ages $>$ 5 Gyr in Figure \ref{fig:proof2}. Unlike the $J_\phi$ result,  the separation between the groups in $J_R$ is similar for stars with ages $<$ 1 Gyr and ages $>$ 5 Gyr. Overall, the mean $J_R$ amplitude increases with stellar age. The $J_z$ result for young stars with ages $<$ 1 Gyr in Figure \ref{fig:proof} shows that $J_z$ decreases for young stars $<$ 1 Gyr with increasing \feh. This is not seen for older stars $>$ 5 Gyr in Figure \ref{fig:proof2}. Similarly to $J_R$, whilst the mean $J_z$ increases for both N=1 and N=2 groups for ages $<$ 1 Gry to ages $>$ 5 Gyr, the mean action separation between the two groups does not significantly change (although the presumably random variation in action difference across \feh\ appears higher for older stars).  We note that the magnitude of the difference in age does not, in general, correspond to the action separation. This is with the exception of the two pairs of stars in groups N=1 and N=2 with $\approx$ 0 age difference at ages $<$ 1 Gyr, for $J_\phi$ and $J_R$. These pairs of stars show no measurable action difference in $J_\phi$ and $J_R$.  The intrinsic dispersion around the mean actions is high (on order of the mean action values themselves for $J_R$ and $J_z$), and significantly larger for older stars. This indicates that just because the stars separate in mean orbital actions at a given metallicity, individual orbital actions themselves will not necessarily be able to be well recovered from the abundance vector for a star \citep[see also][]{Beane2018}.

\section{Discussion}

There has been an immense increase in the size and dimensionality of chemical and dynamical data of stars. In this work, we seek to interpret this data by understanding the relationship between individual abundance enhancements of stars and age.  This is motivated by our understanding of the formation of galaxies, which gives us a picture of stars in our own Milky Way forming in a gas disk. Our understanding of galactic evolution suggests that this gas disk, and hence the radial extent of stellar birth sites, will grow with time.  Our understanding of nucleosynthesis within, and chemical enrichment from those stars, implies that each generation of stars will contain signatures of prior generations \citep{Rix2013, BH2016}. These insights combine into a picture where stars of a given age form from gas which is chemically segregated into mono-abundance rings, whose spatial extent gradually increases over time. Hence, the chemical abundance distributions of stars across the Galaxy are expected to tell us the story of how our Galaxy formed, while their locations or orbits tell us something about where. 

This expectation has been exploited with  forward models of chemical enrichment. These models attempt to capture the physics of galaxy formation and chemical enrichment, given the observed abundance distributions \citep[e.g.][]{V2019, Clarke2019, M2019, Robyn2018, Loebman2016, Nidever2014}. Two key uncertainties limit this approach: the extent to which the stars will lose their memory of the location of their birth sites as they migrate around the Galaxy through secular (e.g. resonances with bars and spiral arms) and non-secular (e.g. interactions with Galactic satellites or gas clouds) dynamical processes; and the sources, timescales and yields of enrichment. A second approach to interpretation is to chemically tag stars to take advantage of the expectation that stars formed in the same gas cloud will have identical chemical signatures \citep{Freeman2002, Krumholz2018}. The main uncertainty here is to what extent star clusters will be chemically distinct within the measurement errors \citep{Ting2015, Liu2019, Bovy2019, Ness2018a}. 

This paper takes a step back from directly re-constructing Galactic history to instead explore where the information is in the data itself. We see from the data that individual element abundances of stars show rich and varied correlations with age, spatial location and orbital actions.  We have examined the relationships between age and abundance for the low-$\alpha$ stars of the disk, using 18 elements measured from the \apogee\ red clump spectra. Our typical measurement precisions are $<$ 0.03 dex and we do not have to consider diffusion and other astrophysical abundance offsets as a function of stellar parameter space \citep[e.g.][]{Liu2019, Souto2019}. We find that most abundances indicate age, albeit some only weakly. The elements C and N have the largest age gradients, presumably as a consequence of mass-dependent dredge up on the giant branch that renders them age indicators. The elements Al, O and Mg are the next most correlated elements with age (although many of these element age-abundance trends are non-linear and flatten and even appear to turn over at large ages). Presumably the age trends seen for Al, Mg and O, are a consequence of Galactic chemical evolution rather than surface abundance changes (as for C and N). 

We measure the intrinsic dispersion around the age-abundance trends for each abundance,  [X/Fe]. That is, the dispersion around the age-abundance relation that is not explained by the measurement errors. We measure this intrinsic dispersion in two ways. First, we consider only the $\approx$ 70 red clump stars at solar metallicity with asteroseismic ages. For these stars, we calculate the abundance dispersion around a 2nd-order polynomial line of best fit to each age-abundance trend and determine what proportion of this dispersion is not explained by the measurement errors. We find that for the low-$\alpha$ stars in the disk, most abundances have small intrinsic dispersion values at a given age at solar \feh, in line with what is found from high fidelity boutique analyses of solar-twin stars  \citep{Bedell2018}. The solar-twin analyses examined about 100 low-$\alpha$ stars in the solar neighbourhood, making 66 abundance measurements from R = 115,000 data with SNR $>$ 100, finding a mean intrinsic dispersion of $\approx$ 0.02 dex and a range of 0 -- 0.06 dex. We find quite consistent results, with a mean intrinsic dispersion of $\approx$ 0.02 dex for our 17 elements, with a range of $\approx$ 0 -- 0.08 dex (with the exception of P which has an intrinsic dispersion of 0.12 dex but which is not measured in the solar-twin sample). 

For our second approach, we use the 600 \apogee\ stars with asteroseismic ages 0.3 $<$ $\tau_{age}$ $<$ 11.7 Gyr, and metallicities --0.5 $<$ \feh\ $<$ 0.45 dex, to model the relationship between the abundance vector and the labels of ($\tau_{age}$, \feh), across the parameter space of these 600 stars. We do this using a 2nd-order polynomial model and calculate the intrinsic scatter around this model for each abundance feature, across age and \feh, given the errors on each of the abundances. This is an effective measure of the mean intrinsic dispersion around this model,  for each element, across metallicities from -0.5 $<$ \feh\ $<$ 0.45 dex and across ages 0.3 $<$ $\tau_{age}$ $<$ 11.7 Gyr (although distributed around a mean age of 3.3 Gyr). The intrinsic scatter from this model should represent an upper bound on the intrinsic dispersion at solar \feh\ and our two measures of the intrinsic dispersion are consistent within the errors for all except three elements:  Cr, S and Co. We report a mean intrinsic scatter across ($\tau_{age}$, \feh) of $\approx$ 0.02 dex for 16 elements, with a range of $\approx$ 0 -- 0.10 dex. The elements Si, Cr, S, Ca Ni Ti and Mg all have scatters $<$ 0.02 dex (i.e. below the mean of our sample). That the $\alpha$-elements have very small scatter values is not surprising, given as we are conditioning on low $\alpha$-enhancement. The element O is anomalous among the $\alpha$-elements with relatively high scatter of 0.04 dex, although we note it is often reported as behaving dissimilar to the other $\alpha$-elements \citep[e.g.][]{Buder2018}. The elements Mn, K and Al all have scatters of $\approx$ 0.02 dex, which is the mean scatter of the measurements. The elements P, Co, C, N, V and O have the largest scatters of $>$ 0.02 dex (i.e. above the mean scatter of our sample), and range from $\approx$ 0.04 -- 0.10 dex. P has an anomalously high  scatter of 0.10 dex (and intrinsic dispersion at solar \feh\ of 0.12 dex). However, this element is also poorly measured, so any inaccuracy in the measurement uncertainty will more dramatically propagate into this determination. We have no reason however to suspect our error estimate on P is incorrect, or an anomalous underestimation, and this element is not available for comparison from the solar-twin study. Furthermore, there are very few studies for this element and the available literature indicates that massive stars produce three times more P than predicted, as no yields can even nearly produce the solar abundance of this element \citep{C2014}.

We find that our model, of stellar ages and \feh\ given abundances, and its corresponding scatter vector, can well generate the abundance vector for our large set of $\approx$ 15,000 low-$\alpha$ red clump stars, that are distributed across the disk.  We find no preferentially poorer fit of the model as a function of Galactic location. Therefore, the model's scatter determination is a measure of the intrinsic dispersion of the abundances around the age-abundance relation for stars distributed across a large Galactic radius, from 4 -- 16 kpc, well beyond the solar neighbourhood. For most elements, the abundances [X/Fe] of Galactic disk stars can be predicted well (on average $\approx$ 0.02 dex), across a wide range of Galactocentric radii, given age and \feh. 

The intrinsic dispersion around the age-abundance relation across age and \feh\ also measures the prospect of distinguishing different sites of star formation at a given time, which may have different abundance signatures. This also therefore represents the precision with which each element must be measured to pursue chemical tagging of individual disk stars \citep{BH2010}. That is, to assign stars to a unique birth site from their vector of element abundances. However, there is also empirical element variation within star formation sites (open clusters) that is on the order of the intrinsic dispersion itself, for many elements \citep{Ness2018a, Bovy2016, Liu2019b}. Therefore, the intrinsic dispersion does not necessarily actually represent the full variance \textit{between} different individual star formation sites.  It is possible to recover simulated clusters from their abundances alone, which presumably links to their different ages, as traced by abundance differences \citep{PJ2019}. A further test will be to see if simulated individual clusters formed at the same time can be recovered, with abundances drawn from their age-relations as shown, for example, in Figures \ref{fig:ageabund2a} and \ref{fig:gradients}. 

Our results, where by most elements have small intrinsic dispersion measurements, do not diminish the value or importance of stellar abundances to reconstruct the formation and evolution history of the Milky Way's disk. On the contrary, all of the trends we see reveal the richness of the abundances to indicate age and orbits, projected across a large spatial extent.  Although we need extremely high precision to reconstruct clusters, we already have enough precision to quite clearly tackle Galactic evolution using the distribution of mean (abundance, age, radius, orbit). The ground breaking \textit{Gaia} mission has shown that the Galactic disk displays numerous signatures of dynamical structure on local and global scales \citep[e.g.][]{Antoja2018, Trick2018, BH2019, Ted2019b}.  The extent of the chemo-dynamical architecture across the disk is presumably linked to numbers of initial disk forming clusters, cluster masses and migratory events of stars across the disk over time. We see detailed chemo-dynamical structure in our data that presumably links to this. We find that a simple clustering of vertically confined stars of the disk using their abundances (at fixed \feh) separates groups of stars into populations of different mean ages and mean orbital actions \citep[see also][]{Beane2018}. We also see the effect of radial migration, projected into the increasing $J_\phi$ separation of pairs of stars grouped by their abundance similarity at fixed \feh, over time.
 
 The significant opportunity and challenge in tracing the formation of the disk given these data, will be as to how to best combine stars in order to measure underlying chemo-dynamical structure to constrain its formation and evolution. The $\approx$ 0.02 dex (on average) precision on individual abundance measurements, that we find is needed to get at any inter-birth cluster element variation using individual stars, is an unattainable aim with current technology for most large surveys. This is particularly so over any span in parameter space in evolutionary state (as note here we use only the red clump to obtain our high precision abundance measurements shown in Figure 1).  However, by combining stars, a precision of $<<$ 0.02 dex on a population basis (i.e. the  mean abundance value for many stars) will be straightforward to achieve.  
 
 Perhaps most importantly, the abundance measurements and abundance-age correlations we report here are strongly constraining with respect to stellar nucleosynthetic yields and chemical evolution models of the low-$\alpha$ disk across its radial extent.  Examining the ratio of elements within nucleosynthetic families (and also between them), for groups of stars across the disk for single age populations, is one powerful approach in order to isolate particular chemical enrichment events and access the corresponding signatures of the star formation history over time \citep[e.g.][]{Weinberg2018, Blancato2019}. Correspondingly, there is a  significant opportunity to use the measurements we have in hand to develop data-driven based approaches to set nucleosynthetic yield tables and in building chemical evolution models from the data itself.

\section{Conclusion}

We have reported the rich and varied relationships between 18 chemical abundances of stars across the low-$\alpha$ disk in \apogee\ and their ages, as a function of \feh, for stars spanning -0.5 $<$ \feh\ $<$ 0.45 dex. The abundances, including those within single nucleosynthetic families (e.g. $\alpha$-elements, light elements, iron-peak elements), have unique trends with age. We find that many individual abundances for stars contain only marginal additional information beyond being indicators of age, at fixed \feh. Therefore, for good fraction of the low-$\alpha$ disk, age and \feh\ alone can predict the other abundances measured by \apogee\ to high precision. Furthermore, this suggests that a star's \feh, age and membership in the high- or low-$\alpha$ sequence are all (or nearly all that) is needed to determine a star's birth radius \citep[][]{Frankel2018}. Subsequently, for the low-$\alpha$ sequence, stars with the same \feh\ and age most likely were born at the same radial annulus. 

Element abundance precisions of $<$ 0.1 dex are a realistic and achievable goal for multi-million star surveys. In the case of \apogee\ quality spectra, this precision is achieved fairly trivially, at SNR $>$ 30 for most elements \citep[e.g.][]{Ness2018, Ting2018}. Precisions of $<$ 0.02 dex, which is the mean intrinsic dispersion we measure around the age-abundance trends, for our set of elements from \apogee\ spectra, is however, not an achievable goal for large surveys, given current stellar models and infrastructure. In the regime of large stellar samples, the lesser and more achievable precision of $<$ 0.1 dex, particularly across a wider parameter space than that considered here with our red clump sample, still affords tremendous opportunity to reconstruct the Milky Way's formation. From the many millions of densely sampled stars from future surveys, such as  \sloanv's \mwm\ \citep{Kollmeier2017}, {\it WEAVE} \citep{Bonifacio2016}, {\it 4MOST} \citep{deJong2019}, {\it GALAH} \citep{deSilva2015}, {\it PFS} \citep{PFS2018}, {\it LAMOST} \citep{Newberg2012} and {\it MOONS} \citep{C2014}, we have tremendous opportunity to work out how to optimally combine stars (e.g. spatially, dynamically, temporally). In doing so, we can expect to reveal the empirical characteristics of the chemo-dynamical structure in the Milky Way across a vast range of scales \citep[e.g.][]{Harshil2019, Xiang2018}. 
 
\section{Acknowledgements} 

We thank the Flatiron astronomy data group for helpful discussions. 

Melissa Ness is in part supported by a Sloan Foundation Fellowship. This work was performed in part during the \textit{Gaia}19 workshop at the Kavli Institute for Theoretical Physics at the University of California, Santa Barbara.

KVJ’s contributions were supported in part by the National Science Foundation under grants NSF PHY-1748958 and NSF AST-1614743. Her work was performed in part during the \textit{Gaia}19 workshop and the 2019 Santa Barbara \textit{Gaia} Sprint (also supported by the Heising-Simons Foundation), both hosted by the Kavli Institute for Theoretical Physics at the University of California, Santa Barbara.

KB is supported by the NSF Graduate Research Fellowship under grant number DGE 16-44869

KH is partially support by a TDA/Scialog Grant  from the Research Cooperation ``A Gaia-Enabled View of Chemical Homogeneity''. KH is also partially support by the Center for Astrophysical Plasma Properties (CAPP) sponsored by the Department of Energy.”

A.B. was supported in part by the Roy \& Diana Vagelos Program in the Molecular Life Sciences and the Roy \& Diana Vagelos Challenge Award.


\begin{thebibliography}{}
\expandafter\ifx\csname natexlab\endcsname\relax\def\natexlab#1{#1}\fi

\bibitem[{{Adibekyan} {et~al.}(2013){Adibekyan}, {Figueira}, {Santos},
  {Hakobyan}, {Sousa}, {Pace}, {Delgado Mena}, {Robin}, {Israelian}, \&
  {Gonz{\'a}lez Hern{\'a}ndez}}]{Adi2013}
{Adibekyan}, V.~Z., {Figueira}, P., {Santos}, N.~C., {et~al.} 2013, \aap, 554,
  A44

\bibitem[{{Antoja} {et~al.}(2018){Antoja}, {Helmi}, {Romero-G{\'o}mez}, {Katz},
  {Babusiaux}, {Drimmel}, {Evans}, {Figueras}, {Poggio}, {Reyl{\'e}}, {Robin},
  {Seabroke}, \& {Soubiran}}]{Antoja2018}
{Antoja}, T., {Helmi}, A., {Romero-G{\'o}mez}, M., {et~al.} 2018, \nat, 561,
  360

\bibitem[{{Armillotta} {et~al.}(2018{\natexlab{a}}){Armillotta}, {Krumholz}, \&
  {Fujimoto}}]{Arm2018}
{Armillotta}, L., {Krumholz}, M.~R., \& {Fujimoto}, Y. 2018{\natexlab{a}},
  \mnras, 481, 5000

\bibitem[{{Armillotta} {et~al.}(2018{\natexlab{b}}){Armillotta}, {Krumholz}, \&
  {Fujimoto}}]{A2018}
---. 2018{\natexlab{b}}, ArXiv e-prints, arXiv:1807.01712

\bibitem[{{Beane} {et~al.}(2018){Beane}, {Ness}, \& {Bedell}}]{Beane2018}
{Beane}, A., {Ness}, M.~K., \& {Bedell}, M. 2018, \apj, 867, 31

\bibitem[{{Bedell} {et~al.}(2018){Bedell}, {Bean}, {Melendez}, {Spina},
  {Ramirez}, {Asplund}, {Alves-Brito}, {dos Santos}, {Dreizler}, {Yong},
  {Monroe}, \& {Casagrande}}]{Bedell2018}
{Bedell}, M., {Bean}, J.~L., {Melendez}, J., {et~al.} 2018, ArXiv e-prints,
  arXiv:1802.02576

\bibitem[{{Bensby} {et~al.}(2012){Bensby}, {Alves-Brito}, {Oey}, {Yong}, \&
  {Mel{\'e}ndez}}]{Bensby2012}
{Bensby}, T., {Alves-Brito}, A., {Oey}, M.~S., {Yong}, D., \& {Mel{\'e}ndez},
  J. 2012, in Astronomical Society of the Pacific Conference Series, Vol. 458,
  Galactic Archaeology: Near-Field Cosmology and the Formation of the Milky
  Way, ed. W.~{Aoki}, M.~{Ishigaki}, T.~{Suda}, T.~{Tsujimoto}, \&
  N.~{Arimoto}, 171

\bibitem[{{Bensby} {et~al.}(2017){Bensby}, {Feltzing}, {Gould}, {Yee},
  {Johnson}, {Asplund}, {Mel{\'e}ndez}, {Lucatello}, {Howes}, {McWilliam},
  {Udalski}, {Szyma{\'n}ski}, {Soszy{\'n}ski}, {Poleski}, {Wyrzykowski},
  {Ulaczyk}, {Koz{\l}owski}, {Pietrukowicz}, {Skowron}, {Mr{\'o}z}, {Pawlak},
  {Abe}, {Asakura}, {Bhattacharya}, {Bond}, {Bennett}, {Hirao}, {Nagakane},
  {Koshimoto}, {Sumi}, {Suzuki}, \& {Tristram}}]{Bensby2017}
{Bensby}, T., {Feltzing}, S., {Gould}, A., {et~al.} 2017, \aap, 605, A89

\bibitem[{{Binney} \& {Tremaine}(2008)}]{BT2008}
{Binney}, J., \& {Tremaine}, S. 2008, {Galactic Dynamics: Second Edition}

\bibitem[{{Blancato}(2018)}]{Blancato2019}
{Blancato}. 2018, arXiv e-prints, arXiv:1810.12325

\bibitem[{{Bland-Hawthorn} \& {Gerhard}(2016)}]{BH2016}
{Bland-Hawthorn}, J., \& {Gerhard}, O. 2016, \araa, 54, 529

\bibitem[{{Bland-Hawthorn} {et~al.}(2010){Bland-Hawthorn}, {Krumholz}, \&
  {Freeman}}]{BH2010}
{Bland-Hawthorn}, J., {Krumholz}, M.~R., \& {Freeman}, K. 2010, \apj, 713, 166

\bibitem[{{Bland-Hawthorn} {et~al.}(2019){Bland-Hawthorn}, {Sharma},
  {Tepper-Garcia}, {Binney}, {Freeman}, {Hayden}, {Kos}, {De Silva}, {Ellis},
  {Lewis}, {Asplund}, {Buder}, {Casey}, {D'Orazi}, {Duong}, {Khanna}, {Lin},
  {Lind}, {Martell}, {Ness}, {Simpson}, {Zucker}, {Zwitter}, {Kafle},
  {Quillen}, {Ting}, \& {Wyse}}]{BH2019}
{Bland-Hawthorn}, J., {Sharma}, S., {Tepper-Garcia}, T., {et~al.} 2019, \mnras,
  arXiv:1809.02658

\bibitem[{{Bonifacio} {et~al.}(2016){Bonifacio}, {Dalton}, {Trager}, {Aguerri},
  {Carrasco}, {Vallenari}, {Abrams}, {Middleton}, \&
  {Say{\`e}de}}]{Bonifacio2016}
{Bonifacio}, P., {Dalton}, G., {Trager}, S., {et~al.} 2016, in SF2A-2016:
  Proceedings of the Annual meeting of the French Society of Astronomy and
  Astrophysics, ed. C.~{Reyl{\'e}}, J.~{Richard}, L.~{Cambr{\'e}sy},
  M.~{Deleuil}, E.~{P{\'e}contal}, L.~{Tresse}, \& I.~{Vauglin}, 267--270

\bibitem[{{Bovy}(2015)}]{Bovy2015}
{Bovy}, J. 2015, ArXiv e-prints, arXiv:1510.06745

\bibitem[{{Bovy}(2016)}]{Bovy2016}
---. 2016, \apj, 817, 49

\bibitem[{{Bovy} {et~al.}(2019){Bovy}, {Leung}, {Hunt}, {Mackereth},
  {Garcia-Hernandez}, \& {Roman-Lopes}}]{Bovy2019}
{Bovy}, J., {Leung}, H.~W., {Hunt}, J. A.~S., {et~al.} 2019, arXiv e-prints,
  arXiv:1905.11404

\bibitem[{{Bovy} {et~al.}(2012){Bovy}, {Rix}, {Liu}, {Hogg}, {Beers}, \&
  {Lee}}]{Bovy2012}
{Bovy}, J., {Rix}, H.-W., {Liu}, C., {et~al.} 2012, \apj, 753, 148

\bibitem[{{Buder} {et~al.}(2018){Buder}, {Asplund}, {Duong}, {Kos}, {Lind},
  {Ness}, {Sharma}, {Bland-Hawthorn}, {Casey}, {De Silva}, {D'Orazi},
  {Freeman}, {Lewis}, {Lin}, {Martell}, {Schlesinger}, {Simpson}, {Zucker},
  {Zwitter}, {Amarsi}, {Anguiano}, {Carollo}, {Casagrande}, {{\v C}otar},
  {Cottrell}, {Da Costa}, {Gao}, {Hayden}, {Horner}, {Ireland}, {Kafle},
  {Munari}, {Nataf}, {Nordlander}, {Stello}, {Ting}, {Traven}, {Watson},
  {Wittenmyer}, {Wyse}, {Yong}, {Zinn}, \& {{\v Z}erjal}}]{Buder2018}
{Buder}, S., {Asplund}, M., {Duong}, L., {et~al.} 2018, \mnras,
  arXiv:1804.06041

\bibitem[{{Cescutti} {et~al.}(2012){Cescutti}, {Matteucci}, {Caffau}, \&
  {Fran{\c c}ois}}]{P2012}
{Cescutti}, G., {Matteucci}, F., {Caffau}, E., \& {Fran{\c c}ois}, P. 2012,
  \aap, 540, A33

\bibitem[{{Cirasuolo} {et~al.}(2014){Cirasuolo}, {Afonso}, {Carollo}, {Flores},
  \& {Maiolino}}]{C2014}
{Cirasuolo}, M., {Afonso}, J., {Carollo}, M., {Flores}, H., \& {Maiolino}, R.
  e.~a. 2014, in PROCSPIE, Vol. 9147, Ground-based and Airborne Instrumentation
  for Astronomy V, 91470N

\bibitem[{{Clarke} {et~al.}(2019){Clarke}, {Debattista}, {Nidever}, {Loebman},
  {Simons}, {Kassin}, {Du}, {Ness}, {Fisher}, {Quinn}, {Wadsley}, {Freeman}, \&
  {Popescu}}]{Clarke2019}
{Clarke}, A.~J., {Debattista}, V.~P., {Nidever}, D.~L., {et~al.} 2019, \mnras,
  484, 3476

\bibitem[{{de Jong} {et~al.}(2019){de Jong}, {Agertz}, {Agudo Berbel}, {Aird},
  {Alexander}, {Amarsi}, {Anders}, {Andrae}, {Ansarinejad}, {Ansorge}, \&
  et~al.}]{deJong2019}
{de Jong}, R.~S., {Agertz}, O., {Agudo Berbel}, A., {et~al.} 2019, arXiv
  e-prints, arXiv:1903.02464

\bibitem[{{De Silva} {et~al.}(2015){De Silva}, {Freeman}, {Bland-Hawthorn},
  {Martell}, {de Boer}, {Asplund}, {Keller}, {Sharma}, {Zucker}, {Zwitter},
  {Anguiano}, {Bacigalupo}, {Bayliss}, {Beavis}, {Bergemann}, {Campbell},
  {Cannon}, {Carollo}, {Casagrande}, {Casey}, {Da Costa}, {D'Orazi}, {Dotter},
  {Duong}, {Heger}, {Ireland}, {Kafle}, {Kos}, {Lattanzio}, {Lewis}, {Lin},
  {Lind}, {Munari}, {Nataf}, {O'Toole}, {Parker}, {Reid}, {Schlesinger},
  {Sheinis}, {Simpson}, {Stello}, {Ting}, {Traven}, {Watson}, {Wittenmyer},
  {Yong}, \& {{\v Z}erjal}}]{deSilva2015}
{De Silva}, G.~M., {Freeman}, K.~C., {Bland-Hawthorn}, J., {et~al.} 2015,
  \mnras, 449, 2604

\bibitem[{{Duong} {et~al.}(2018){Duong}, {Freeman}, {Asplund}, {Casagrande},
  {Buder}, {Lind}, {Ness}, {Bland-Hawthorn}, {De Silva}, {D'Orazi}, {Kos},
  {Lewis}, {Lin}, {Martell}, {Schlesinger}, {Sharma}, {Simpson}, {Zucker},
  {Zwitter}, {Anguiano}, {Da Costa}, {Hyde}, {Horner}, {Kafle}, {Nataf},
  {Reid}, {Stello}, {Ting}, \& {Wyse}}]{Duong2018}
{Duong}, L., {Freeman}, K.~C., {Asplund}, M., {et~al.} 2018, \mnras, 476, 5216

\bibitem[{{Frankel} {et~al.}(2018){Frankel}, {Rix}, {Ting}, {Ness}, \&
  {Hogg}}]{Frankel2018}
{Frankel}, N., {Rix}, H.-W., {Ting}, Y.-S., {Ness}, M.~K., \& {Hogg}, D.~W.
  2018, ArXiv e-prints, arXiv:1805.09198

\bibitem[{{Freeman} \& {Bland-Hawthorn}(2002)}]{Freeman2002}
{Freeman}, K., \& {Bland-Hawthorn}, J. 2002, \araa, 40, 487

\bibitem[{{Gadotti} {et~al.}(2019){Gadotti}, {S{\'a}nchez-Bl{\'a}zquez},
  {Falc{\'o}n-Barroso}, {Husemann}, {Seidel}, {P{\'e}rez}, {de
  Lorenzo-C{\'a}ceres}, {Martinez-Valpuesta}, {Fragkoudi}, {Leung}, {van de
  Ven}, {Leaman}, {Coelho}, {Martig}, {Kim}, {Neumann}, \&
  {Querejeta}}]{Gadotti2019}
{Gadotti}, D.~A., {S{\'a}nchez-Bl{\'a}zquez}, P., {Falc{\'o}n-Barroso}, J.,
  {et~al.} 2019, \mnras, 482, 506

\bibitem[{{Gaia Collaboration}(2018)}]{Gaia2018}
{Gaia Collaboration}. 2018, VizieR Online Data Catalog, 1345

\bibitem[{{Gandhi} \& {Ness}(2019)}]{Gandhi2019}
{Gandhi}, S.~S., \& {Ness}, M.~K. 2019, arXiv e-prints, arXiv:1903.04030

\bibitem[{{Genovali} {et~al.}(2014){Genovali}, {Lemasle}, {Bono}, {Romaniello},
  {Fabrizio}, {Ferraro}, {Iannicola}, {Laney}, {Nonino}, {Bergemann},
  {Buonanno}, {Fran{\c c}ois}, {Inno}, {Kudritzki}, {Matsunaga}, {Pedicelli},
  {Primas}, \& {Th{\'e}venin}}]{G2014}
{Genovali}, K., {Lemasle}, B., {Bono}, G., {et~al.} 2014, \aap, 566, A37

\bibitem[{{Gilmore} {et~al.}(2012){Gilmore}, {Randich}, {Asplund}, {Binney},
  {Bonifacio}, {Drew}, {Feltzing}, {Ferguson}, {Jeffries}, {Micela}, \&
  et~al.}]{Gilmore2012}
{Gilmore}, G., {Randich}, S., {Asplund}, M., {et~al.} 2012, The Messenger, 147,
  25

\bibitem[{{Hawkins} {et~al.}(2017){Hawkins}, {Leistedt}, {Bovy}, \&
  {Hogg}}]{Hawkins2017}
{Hawkins}, K., {Leistedt}, B., {Bovy}, J., \& {Hogg}, D.~W. 2017, \mnras, 471,
  722

\bibitem[{{Hawkins} {et~al.}(2018){Hawkins}, {Ting}, \&
  {Walter-Rix}}]{Hawkins2018}
{Hawkins}, K., {Ting}, Y.-S., \& {Walter-Rix}, H. 2018, \apj, 853, 20

\bibitem[{{Hayden} {et~al.}(2015){Hayden}, {Bovy}, {Holtzman}, {Nidever},
  {Bird}, {Weinberg}, {Andrews}, {Majewski}, {Allende Prieto}, {Anders},
  {Beers}, {Bizyaev}, {Chiappini}, {Cunha}, {Frinchaboy},
  {Garc{\'{\i}}a-Her{\'n}andez}, {Garc{\'{\i}}a P{\'e}rez}, {Girardi},
  {Harding}, {Hearty}, {Johnson}, {M{\'e}sz{\'a}ros}, {Minchev}, {O'Connell},
  {Pan}, {Robin}, {Schiavon}, {Schneider}, {Schultheis}, {Shetrone},
  {Skrutskie}, {Steinmetz}, {Smith}, {Wilson}, {Zamora}, \&
  {Zasowski}}]{Hayden2015}
{Hayden}, M.~R., {Bovy}, J., {Holtzman}, J.~A., {et~al.} 2015, \apj, 808, 132

\bibitem[{{Haywood} {et~al.}(2013){Haywood}, {Di Matteo}, {Lehnert}, {Katz}, \&
  {G{\'o}mez}}]{Haywood2013}
{Haywood}, M., {Di Matteo}, P., {Lehnert}, M.~D., {Katz}, D., \& {G{\'o}mez},
  A. 2013, \aap, 560, A109

\bibitem[{{Ho} {et~al.}(2017){Ho}, {Rix}, {Ness}, {Hogg}, {Liu}, \&
  {Ting}}]{Ho2017b}
{Ho}, A.~Y.~Q., {Rix}, H.-W., {Ness}, M.~K., {et~al.} 2017, \apj, 841, 40

\bibitem[{{Hogg} {et~al.}(2016){Hogg}, {Casey}, {Ness}, {Rix},
  {Foreman-Mackey}, {Hasselquist}, {Ho}, {Holtzman}, {Majewski}, {Martell},
  {M{\'e}sz{\'a}ros}, {Nidever}, \& {Shetrone}}]{Hogg2016}
{Hogg}, D.~W., {Casey}, A.~R., {Ness}, M., {et~al.} 2016, \apj, 833, 262

\bibitem[{{Kamdar} {et~al.}(2019){Kamdar}, {Conroy}, {Ting}, {Bonaca},
  {Johnson}, \& {Cargile}}]{Harshil2019}
{Kamdar}, H., {Conroy}, C., {Ting}, Y.-S., {et~al.} 2019, arXiv e-prints,
  arXiv:1902.10719

\bibitem[{{Kollmeier} {et~al.}(2017){Kollmeier}, {Zasowski}, {Rix}, {Johns},
  {Anderson}, {Drory}, {Johnson}, {Pogge}, {Bird}, {Blanc}, {Brownstein},
  {Crane}, {De Lee}, {Klaene}, {Kreckel}, {MacDonald}, {Merloni}, {Ness},
  {O'Brien}, {Sanchez-Gallego}, {Sayres}, {Shen}, {Thakar}, {Tkachenko},
  {Aerts}, {Blanton}, {Eisenstein}, {Holtzman}, {Maoz}, {Nandra}, {Rockosi},
  {Weinberg}, {Bovy}, {Casey}, {Chaname}, {Clerc}, {Conroy}, {Eracleous},
  {G{\"a}nsicke}, {Hekker}, {Horne}, {Kauffmann}, {McQuinn}, {Pellegrini},
  {Schinnerer}, {Schlafly}, {Schwope}, {Seibert}, {Teske}, \& {van
  Saders}}]{Kollmeier2017}
{Kollmeier}, J.~A., {Zasowski}, G., {Rix}, H.-W., {et~al.} 2017, ArXiv
  e-prints, arXiv:1711.03234

\bibitem[{{Krumholz} {et~al.}(2018){Krumholz}, {McKee}, \& {Bland
  -Hawthorn}}]{Krumholz2018}
{Krumholz}, M.~R., {McKee}, C.~F., \& {Bland -Hawthorn}, J. 2018, arXiv
  e-prints, arXiv:1812.01615

\bibitem[{{Leung} \& {Bovy}(2019)}]{Leung2019}
{Leung}, H.~W., \& {Bovy}, J. 2019, arXiv e-prints, arXiv:1902.08634

\bibitem[{{Liu} {et~al.}(2019{\natexlab{a}}){Liu}, {Asplund}, {Yong},
  {Feltzing}, {Dotter}, {Mel{\'e}ndez}, \& {Ram{\'{\i}}rez}}]{Liu2019}
{Liu}, F., {Asplund}, M., {Yong}, D., {et~al.} 2019{\natexlab{a}}, arXiv
  e-prints, arXiv:1902.11008

\bibitem[{{Liu} {et~al.}(2019{\natexlab{b}}){Liu}, {Asplund}, {Yong},
  {Feltzing}, {Dotter}, {Mel{\'e}ndez}, \& {Ram{\'{\i}}rez}}]{Liu2019b}
---. 2019{\natexlab{b}}, arXiv e-prints, arXiv:1902.11008

\bibitem[{{Loebman} {et~al.}(2016){Loebman}, {Debattista}, {Nidever}, {Hayden},
  {Holtzman}, {Clarke}, {Ro{\v{s}}kar}, \& {Valluri}}]{Loebman2016}
{Loebman}, S.~R., {Debattista}, V.~P., {Nidever}, D.~L., {et~al.} 2016, \apj,
  818, L6

\bibitem[{{Luck} \& {Lambert}(2011)}]{Luck2011}
{Luck}, R.~E., \& {Lambert}, D.~L. 2011, \aj, 142, 136

\bibitem[{{Mackereth} {et~al.}(2019{\natexlab{a}}){Mackereth}, {Bovy}, {Leung},
  {Schiavon}, {Trick}, {Chaplin}, {Cunha}, {Feuillet}, {Majewski}, {Martig},
  {Miglio}, {Nidever}, {Pinsonneault}, {Silva Aguirre}, {Sobeck}, {Tayar}, \&
  {Zasowski}}]{Ted2019a}
{Mackereth}, J.~T., {Bovy}, J., {Leung}, H.~W., {et~al.} 2019{\natexlab{a}},
  arXiv e-prints, arXiv:1901.04502

\bibitem[{{Mackereth} {et~al.}(2019{\natexlab{b}}){Mackereth}, {Schiavon},
  {Pfeffer}, {Hayes}, {Bovy}, {Anguiano}, {Allende Prieto}, {Hasselquist},
  {Holtzman}, {Johnson}, {Majewski}, {O'Connell}, {Shetrone}, {Tissera}, \&
  {Fern{\'a}ndez-Trincado}}]{Ted2019b}
{Mackereth}, J.~T., {Schiavon}, R.~P., {Pfeffer}, J., {et~al.}
  2019{\natexlab{b}}, \mnras, 482, 3426

\bibitem[{{Majewski} {et~al.}(2017){Majewski}, {Schiavon}, {Frinchaboy},
  {Allende Prieto}, {Barkhouser}, {Bizyaev}, {Blank}, {Brunner}, {Burton},
  {Carrera}, {Chojnowski}, {Cunha}, {Epstein}, {Fitzgerald}, {Garc{\'{\i}}a
  P{\'e}rez}, {Hearty}, {Henderson}, {Holtzman}, {Johnson}, {Lam}, {Lawler},
  {Maseman}, {M{\'e}sz{\'a}ros}, {Nelson}, {Nguyen}, {Nidever}, {Pinsonneault},
  {Shetrone}, {Smee}, {Smith}, {Stolberg}, {Skrutskie}, {Walker}, {Wilson},
  {Zasowski}, {Anders}, {Basu}, {Beland}, {Blanton}, {Bovy}, {Brownstein},
  {Carlberg}, {Chaplin}, {Chiappini}, {Eisenstein}, {Elsworth}, {Feuillet},
  {Fleming}, {Galbraith-Frew}, {Garc{\'{\i}}a}, {Garc{\'{\i}}a-Hern{\'a}ndez},
  {Gillespie}, {Girardi}, {Gunn}, {Hasselquist}, {Hayden}, {Hekker}, {Ivans},
  {Kinemuchi}, {Klaene}, {Mahadevan}, {Mathur}, {Mosser}, {Muna}, {Munn},
  {Nichol}, {O'Connell}, {Parejko}, {Robin}, {Rocha-Pinto}, {Schultheis},
  {Serenelli}, {Shane}, {Silva Aguirre}, {Sobeck}, {Thompson}, {Troup},
  {Weinberg}, \& {Zamora}}]{Majewski2017}
{Majewski}, S.~R., {Schiavon}, R.~P., {Frinchaboy}, P.~M., {et~al.} 2017, \aj,
  154, 94

\bibitem[{{Martig} {et~al.}(2016){Martig}, {Fouesneau}, {Rix}, {Ness},
  {M{\'e}sz{\'a}ros}, {Garc{\'{\i}}a-Hern{\'a}ndez}, {Pinsonneault},
  {Serenelli}, {Silva Aguirre}, \& {Zamora}}]{Martig2016}
{Martig}, M., {Fouesneau}, M., {Rix}, H.-W., {et~al.} 2016, \mnras, 456, 3655

\bibitem[{{Masseron} \& {Gilmore}(2015)}]{Masseron2015}
{Masseron}, T., \& {Gilmore}, G. 2015, \mnras, 453, 1855

\bibitem[{{McMillan}(2017)}]{Paul2017}
{McMillan}, P.~J. 2017, \mnras, 465, 76

\bibitem[{{Minchev} {et~al.}(2011){Minchev}, {Famaey}, {Combes}, {Di Matteo},
  {Mouhcine}, \& {Wozniak}}]{Minchev2011}
{Minchev}, I., {Famaey}, B., {Combes}, F., {et~al.} 2011, \aap, 527, A147

\bibitem[{{Moll{\'a}} {et~al.}(2019){Moll{\'a}}, {D{\'\i}az}, {Cavichia},
  {Gibson}, {Maciel}, {Costa}, {Ascasibar}, \& {Few}}]{M2019}
{Moll{\'a}}, M., {D{\'\i}az}, {\'A}.~I., {Cavichia}, O., {et~al.} 2019, \mnras,
  482, 3071

\bibitem[{{Ness}(2018)}]{Ness2018}
{Ness}, M. 2018, \pasa, 35, e003

\bibitem[{{Ness} {et~al.}(2015){Ness}, {Hogg}, {Rix}, {Ho}, \&
  {Zasowski}}]{Ness2015}
{Ness}, M., {Hogg}, D.~W., {Rix}, H.-W., {Ho}, A.~Y.~Q., \& {Zasowski}, G.
  2015, \apj, 808, 16

\bibitem[{{Ness} {et~al.}(2016){Ness}, {Hogg}, {Rix}, {Martig}, {Pinsonneault},
  \& {Ho}}]{Ness2016}
{Ness}, M., {Hogg}, D.~W., {Rix}, H.-W., {et~al.} 2016, \apj, 823, 114

\bibitem[{{Ness} {et~al.}(2013){Ness}, {Freeman}, {Athanassoula},
  {Wylie-de-Boer}, {Bland-Hawthorn}, {Asplund}, {Lewis}, {Yong}, {Lane}, \&
  {Kiss}}]{Ness2013a}
{Ness}, M., {Freeman}, K., {Athanassoula}, E., {et~al.} 2013, \mnras, 430, 836

\bibitem[{{Ness} {et~al.}(2018){Ness}, {Rix}, {Hogg}, {Casey}, {Holtzman},
  {Fouesneau}, {Zasowski}, {Geisler}, {Shetrone}, {Minniti}, {Frinchaboy}, \&
  {Roman-Lopes}}]{Ness2018a}
{Ness}, M., {Rix}, H.-W., {Hogg}, D.~W., {et~al.} 2018, \apj, 853, 198

\bibitem[{{Newberg} {et~al.}(2012){Newberg}, {Carlin}, {Chen}, {Deng},
  {L{\'e}pine}, {Liu}, {Yang}, {Yuan}, {Zhang}, {Zhang}, {Legue Working Group},
  \& {Lamost-Plus Partnership}}]{Newberg2012}
{Newberg}, H.~J., {Carlin}, J.~L., {Chen}, L., {et~al.} 2012, in Astronomical
  Society of the Pacific Conference Series, Vol. 458, Galactic Archaeology:
  Near-Field Cosmology and the Formation of the Milky Way, ed. W.~{Aoki},
  M.~{Ishigaki}, T.~{Suda}, T.~{Tsujimoto}, \& N.~{Arimoto}, 405

\bibitem[{{Nidever} {et~al.}(2014){Nidever}, {Bovy}, {Bird}, {Andrews},
  {Hayden}, {Holtzman}, {Majewski}, {Smith}, {Robin}, {Garc{\'{\i}}a
  P{\'e}rez}, {Cunha}, {Allende Prieto}, {Zasowski}, {Schiavon}, {Johnson},
  {Weinberg}, {Feuillet}, {Schneider}, {Shetrone}, {Sobeck},
  {Garc{\'{\i}}a-Hern{\'a}ndez}, {Zamora}, {Rix}, {Beers}, {Wilson},
  {O'Connell}, {Minchev}, {Chiappini}, {Anders}, {Bizyaev}, {Brewington},
  {Ebelke}, {Frinchaboy}, {Ge}, {Kinemuchi}, {Malanushenko}, {Malanushenko},
  {Marchante}, {M{\'e}sz{\'a}ros}, {Oravetz}, {Pan}, {Simmons}, \&
  {Skrutskie}}]{Nidever2014}
{Nidever}, D.~L., {Bovy}, J., {Bird}, J.~C., {et~al.} 2014, \apj, 796, 38

\bibitem[{{Pinsonneault} {et~al.}(2018){Pinsonneault}, {Elsworth}, {Tayar},
  {Serenelli}, {Stello}, {Zinn}, {Mathur}, {Garc{\'{\i}}a}, {Johnson},
  {Hekker}, {Huber}, {Kallinger}, {M{\'e}sz{\'a}ros}, {Mosser}, {Stassun},
  {Girardi}, {Rodrigues}, {Silva Aguirre}, {An}, {Basu}, {Chaplin}, {Corsaro},
  {Cunha}, {Garc{\'{\i}}a-Hern{\'a}ndez}, {Holtzman}, {J{\"o}nsson},
  {Shetrone}, {Smith}, {Sobeck}, {Stringfellow}, {Zamora}, {Beers},
  {Fern{\'a}ndez-Trincado}, {Frinchaboy}, {Hearty}, \& {Nitschelm}}]{Pins2018}
{Pinsonneault}, M.~H., {Elsworth}, Y.~P., {Tayar}, J., {et~al.} 2018, ArXiv
  e-prints, arXiv:1804.09983

\bibitem[{{Price-Jones} \& {Bovy}(2019)}]{PJ2019}
{Price-Jones}, N., \& {Bovy}, J. 2019, arXiv e-prints, arXiv:1902.08201

\bibitem[{{Rix} \& {Bovy}(2013)}]{Rix2013}
{Rix}, H.-W., \& {Bovy}, J. 2013, \aapr, 21, 61

\bibitem[{{Ro{\v s}kar} {et~al.}(2008){Ro{\v s}kar}, {Debattista}, {Quinn},
  {Stinson}, \& {Wadsley}}]{Roskar2008}
{Ro{\v s}kar}, R., {Debattista}, V.~P., {Quinn}, T.~R., {Stinson}, G.~S., \&
  {Wadsley}, J. 2008, \apjl, 684, L79

\bibitem[{{Sanderson} {et~al.}(2018){Sanderson}, {Wetzel}, {Loebman}, {Sharma},
  {Hopkins}, {Garrison-Kimmel}, {Faucher-Gigu{\`e}re}, {Kere{\v{s}}}, \&
  {Quataert}}]{Robyn2018}
{Sanderson}, R.~E., {Wetzel}, A., {Loebman}, S., {et~al.} 2018, arXiv e-prints,
  arXiv:1806.10564

\bibitem[{{Schlesinger} {et~al.}(2014){Schlesinger}, {Johnson}, {Rockosi},
  {Lee}, {Beers}, {Harding}, {Allende Prieto}, {Bird}, {Sch{\"o}nrich},
  {Yanny}, {Schneider}, {Weaver}, \& {Brinkmann}}]{Katie2014}
{Schlesinger}, K.~J., {Johnson}, J.~A., {Rockosi}, C.~M., {et~al.} 2014, \apj,
  791, 112

\bibitem[{{Sellwood} \& {Binney}(2002)}]{Selwood2002}
{Sellwood}, J.~A., \& {Binney}, J.~J. 2002, \mnras, 336, 785

\bibitem[{{Silva Aguirre} {et~al.}(2018){Silva Aguirre}, {Bojsen-Hansen},
  {Slumstrup}, {Casagrande}, {Kawata}, {Ciuc{\v a}}, {Handberg}, {Lund},
  {Mosumgaard}, {Huber}, {Johnson}, {Pinsonneault}, {Serenelli}, {Stello},
  {Tayar}, {Bird}, {Cassisi}, {Hon}, {Martig}, {Nissen}, {Rix},
  {Sch{\"o}nrich}, {Sahlholdt}, {Trick}, \& {Yu}}]{VSA2018}
{Silva Aguirre}, V., {Bojsen-Hansen}, M., {Slumstrup}, D., {et~al.} 2018,
  \mnras, 475, 5487

\bibitem[{{Souto} {et~al.}(2019){Souto}, {Allende Prieto}, {Cunha},
  {Pinsonneault}, {Smith}, {Garcia-Dias}, {Bovy}, {Garcia-Hernandez},
  {Holtzman}, {Johnson}, {Jonsson}, {Majewski}, {Shetrone}, {Sobeck}, {Zamora},
  {Pan}, \& {Nitschelm}}]{Souto2019}
{Souto}, D., {Allende Prieto}, C., {Cunha}, K., {et~al.} 2019, arXiv e-prints,
  arXiv:1902.10199

\bibitem[{{Steinmetz} {et~al.}(2006){Steinmetz}, {Zwitter}, {Siebert},
  {Watson}, {Freeman}, {Munari}, {Campbell}, {Williams}, {Seabroke}, {Wyse},
  {Parker}, {Bienaym{\'e}}, {Roeser}, {Gibson}, {Gilmore}, {Grebel}, {Helmi},
  {Navarro}, {Burton}, {Cass}, {Dawe}, {Fiegert}, {Hartley}, {Russell},
  {Saunders}, {Enke}, {Bailin}, {Binney}, {Bland-Hawthorn}, {Boeche}, {Dehnen},
  {Eisenstein}, {Evans}, {Fiorucci}, {Fulbright}, {Gerhard}, {Jauregi}, {Kelz},
  {Mijovi{\'c}}, {Minchev}, {Parmentier}, {Pe{\~n}arrubia}, {Quillen}, {Read},
  {Ruchti}, {Scholz}, {Siviero}, {Smith}, {Sordo}, {Veltz}, {Vidrih}, {von
  Berlepsch}, {Boyle}, \& {Schilbach}}]{Steinmetz2006}
{Steinmetz}, M., {Zwitter}, T., {Siebert}, A., {et~al.} 2006, \aj, 132, 1645

\bibitem[{{Tamura} {et~al.}(2018){Tamura}, {Takato}, {Shimono}, {Moritani},
  {Yabe}, {Ishizuka}, {Kamata}, {Ueda}, {Aghazarian}, {Arnouts}, {Barkhouser},
  {Balard}, {Barette}, {Belhadi}, {Burnham}, {Caplar}, {Carr}, {Chabaud},
  {Chang}, {Chen}, {Chou}, {Chu}, {Cohen}, {de Almeida}, {de Oliveira}, {de
  Oliveira}, {Dekany}, {Dohlen}, {dos Santos}, {dos Santos}, {Ellis},
  {Fabricius}, {Ferreira}, {Furusawa}, {Garcia-Carpio}, {Golebiowski}, {Gross},
  {Gunn}, {Hammond}, {Harding}, {Hart}, {Heckman}, {Ho}, {Hope}, {Hover},
  {Hsu}, {Hu}, {Huang}, {Jamal}, {Jaquet}, {Jeschke}, {Jing}, {Kado-Fong},
  {Karr}, {Kimura}, {King}, {Koike}, {Komatsu}, {Le Brun}, {Le F{\`e}vre}, {Le
  Fur}, {Le Mignant}, {Ling}, {Loomis}, {Lupton}, {Madec}, {Mao}, {Marchesini},
  {Marrara}, {Medvedev}, {Mineo}, {Minowa}, {Murayama}, {Murray}, {Ohyama},
  {Onodera}, {Orndorff}, {Pascal}, {Peebles}, {Pernot}, {Pourcelot}, {Reiley},
  {Reinecke}, {Roberts}, {Rosa}, {Rousselle}, {Schmitt}, {Schwochert},
  {Seiffert}, {Siddiqui}, {Smee}, {Sodr{\'e}}, {Steinkraus}, {Strauss},
  {Surace}, {Tait}, {Takada}, {Tamura}, {Tanaka}, {Tanaka}, {Thakar},
  {Verducci}, {Vibert}, {Wang}, {Wang}, {Wen}, {Werner}, {Yamada}, {Yan},
  {Yasuda}, {Yoshida}, \& {Yoshida}}]{PFS2018}
{Tamura}, N., {Takato}, N., {Shimono}, A., {et~al.} 2018, in Society of
  Photo-Optical Instrumentation Engineers (SPIE) Conference Series, Vol. 10702,
  Ground-based and Airborne Instrumentation for Astronomy VII, 107021C

\bibitem[{{Ting} {et~al.}(2015){Ting}, {Conroy}, \& {Goodman}}]{Ting2015}
{Ting}, Y.-S., {Conroy}, C., \& {Goodman}, A. 2015, \apj, 807, 104

\bibitem[{{Ting} {et~al.}(2012){Ting}, {Freeman}, {Kobayashi}, {De Silva}, \&
  {Bland-Hawthorn}}]{Ting2012}
{Ting}, Y.-S., {Freeman}, K.~C., {Kobayashi}, C., {De Silva}, G.~M., \&
  {Bland-Hawthorn}, J. 2012, \mnras, 421, 1231

\bibitem[{{Ting} {et~al.}(2018){Ting}, {Hawkins}, \& {Rix}}]{Ting2018}
{Ting}, Y.-S., {Hawkins}, K., \& {Rix}, H.-W. 2018, \apjl, 858, L7

\bibitem[{{Trick} {et~al.}(2018){Trick}, {Coronado}, \& {Rix}}]{Trick2018}
{Trick}, W.~H., {Coronado}, J., \& {Rix}, H.-W. 2018, ArXiv e-prints,
  arXiv:1805.03653

\bibitem[{{Valentini} {et~al.}(2019){Valentini}, {Borgani}, {Bressan},
  {Murante}, {Tornatore}, \& {Monaco}}]{V2019}
{Valentini}, M., {Borgani}, S., {Bressan}, A., {et~al.} 2019, \mnras, 485, 1384

\bibitem[{{Weinberg} {et~al.}(2018){Weinberg}, {Holtzman}, {Hasselquist},
  {Bird}, {Johnson}, {Shetrone}, {Sobeck}, {Allende Prieto}, {Bizyaev},
  {Carrera}, {Cohen}, {Cunha}, {Ebelke}, {Fernandez-Trincado},
  {Garcia-Hernandez}, {Hayes}, {Jonsson}, {Lane}, {Majewski}, {Malanushenko},
  {Meszaros}, {Nidever}, {Nitschelm}, {Pan}, {Schiavon}, {Schneider}, {Wilson},
  \& {Zamora}}]{Weinberg2018}
{Weinberg}, D.~H., {Holtzman}, J.~A., {Hasselquist}, S., {et~al.} 2018, arXiv
  e-prints, arXiv:1810.12325

\bibitem[{{Xiang} {et~al.}(2018){Xiang}, {Shi}, {Liu}, {Yuan}, {Chen}, {Huang},
  {Wang}, {Wu}, {Tian}, {Huo}, {Zhang}, \& {Zhang}}]{Xiang2018}
{Xiang}, M., {Shi}, J., {Liu}, X., {et~al.} 2018, \apjs, 237, 33

\bibitem[{{Yanny} {et~al.}(2009){Yanny}, {Rockosi}, {Newberg}, {Knapp},
  {Adelman-McCarthy}, {Alcorn}, {Allam}, {Allende Prieto}, {An}, {Anderson},
  {Anderson}, {Bailer-Jones}, {Bastian}, {Beers}, {Bell}, {Belokurov},
  {Bizyaev}, {Blythe}, {Bochanski}, {Boroski}, {Brinchmann}, {Brinkmann},
  {Brewington}, {Carey}, {Cudworth}, {Evans}, {Evans}, {Gates}, {G{\"a}nsicke},
  {Gillespie}, {Gilmore}, {Nebot Gomez-Moran}, {Grebel}, {Greenwell}, {Gunn},
  {Jordan}, {Jordan}, {Harding}, {Harris}, {Hendry}, {Holder}, {Ivans},
  {Ivezi{\v c}}, {Jester}, {Johnson}, {Kent}, {Kleinman}, {Kniazev},
  {Krzesinski}, {Kron}, {Kuropatkin}, {Lebedeva}, {Lee}, {French Leger},
  {L{\'e}pine}, {Levine}, {Lin}, {Long}, {Loomis}, {Lupton}, {Malanushenko},
  {Malanushenko}, {Margon}, {Martinez-Delgado}, {McGehee}, {Monet}, {Morrison},
  {Munn}, {Neilsen}, {Nitta}, {Norris}, {Oravetz}, {Owen}, {Padmanabhan},
  {Pan}, {Peterson}, {Pier}, {Platson}, {Re Fiorentin}, {Richards}, {Rix},
  {Schlegel}, {Schneider}, {Schreiber}, {Schwope}, {Sibley}, {Simmons},
  {Snedden}, {Allyn Smith}, {Stark}, {Stauffer}, {Steinmetz}, {Stoughton},
  {SubbaRao}, {Szalay}, {Szkody}, {Thakar}, {Sivarani}, {Tucker}, {Uomoto},
  {Vanden Berk}, {Vidrih}, {Wadadekar}, {Watters}, {Wilhelm}, {Wyse}, {Yarger},
  \& {Zucker}}]{Yanny2009}
{Yanny}, B., {Rockosi}, C., {Newberg}, H.~J., {et~al.} 2009, \aj, 137, 4377

\end{thebibliography}

\end{document}